\documentclass[referee]{aa} 
\usepackage{graphicx}
\usepackage{txfonts}
\usepackage{natbib}
\bibpunct{(}{)}{;}{a}{}{,}
\usepackage[figuresright]{rotating}
\usepackage{aalongtable}
\usepackage{aalongtable,lscape}
\begin{document}
  \title{Upper limits on the solar-neutron flux at the Yangbajing neutron monitor from BATSE-detected solar flares}

  \author{H. Tsuchiya \inst{1}
    \and
    H. Miyasaka \inst{2}
    \and
    E. Takahashi \inst{1}
    \and
    S. Shimoda \inst{1}
    \and
    Y. Yamada \inst{1}
    \and
    I. Kondo \inst{1}
    \and
    K. Makishima \inst{1,3}
    \and
    F. Zhu \inst{4}
    \and
    Y. Tan  \inst{4}
    \and
    H. Hu   \inst{4}
    \and
    Y. Tang \inst{5}
    \and
    J. Zhang \inst{4}
    \and
    H. Lu \inst{4}
    \and
    X. Meng \inst{6}
  }

  \offprints{H. Tsuchiya}

  \institute{RIKEN, 2-1, Hirosawa, Wako, Staitama 351-0198, Japan\\
    \email{harufumi@crab.riken.jp}
    \and
    Caltech, Pasadena, CA 91125, USA\\
    \and
    Department of Physics, University of Tokyo, 7-3-1, Hongo, Bunkyo-ku, Tokyo 113-0033, Japan
    \and
    Institute of High Energy Physics, Beijing, 100049, China \\
    \and
    China Meteorological Administration, Beijing 100049, China
    \and
    Tibet University, Lhasa 850000
  }
  
  \date{Received November 3, 2006; accepted March 12, 2007}

  \abstract
   {} 
   {
     The purpose of this work is to search the Yangbajing neutron monitor data obtained 
     between 1998 October and 2000 June for solar neutrons associated with solar flares.}
   {
     Using the onset times of 166 BATSE-detected flares with the 
     GOES peak flux (1 -- 8~\AA) higher than $1.0 \times 10^{-5}$ $\mathrm{Wm^{-2}}$,
     we prepare for each flare a light curve of the Yangbajing neutron monitor, spanning $\pm$ 1.5 hours 
     from the BATSE onset time. Based on the light curves, 
     a systematic search for solar neutrons in energies above 100~MeV from the 166 flares
     was performed.
}
   { 
     No statistically significant signals due to solar neutrons were found in the present work.
     Therefore, we put upper limits on the $>$ 100~MeV solar-neutron flux for 18 events 
     consisting of 2 X and 16 M class flares. The calculation assumed a power-law
     shaped neutron energy spectrum and three types of neutron emission profiles at the Sun.
     Compared with the other positive neutron detections associated with X-class flares, 
     typical 95\% confidence level upper limits for the two X-class flares 
     are found to be comparable to the lowest and second lowest neutron fluxes at the top of the atmosphere.
     In addition, the upper limits for M-class flares scatter in the range of $10^{-2}$ to 1 neutrons $\mathrm{cm^{-2}s^{-1}}$.
     This provides the first upper limits on the solar-neutron flux from M-class solar flares, 
     using space observatories as well as ground-based neutron monitors.
   }
   {}
   \titlerunning{Upper limits on the solar neutron flux with a neutron monitor}
   \authorrunning{Tsuchiya et al.}
   \maketitle
%

\section{Introduction}
Solar neutrons as well as nuclear gamma rays have been widely used as a good probe 
to the still unclear fundamental mechanisms of ion acceleration associated with solar flares. 
These neutral secondaries emanate deep in the chromosphere via nuclear processes 
between accelerated ions and the solar ambient plasma. 

Early quantitative studies~\citep{HESS1962,L1965a,L1965b,L1969} predicted that 
large solar flares would produce a measurable flux of 
solar neutrons at the Earth, $\sim$ 10 $\mathrm{cm^{-2}s^{-1}}$,
in energies above 10~MeV.
Before the solar cycle 21, various balloon-borne experiments 
attempted to detect the predicted neutrons originating from solar flares~\citep{A1966,D1967,HK1967,D1969,FC1969},
but none gave convincing results.

In the solar cycle 21 and 22, the predicted solar neutrons have actually been detected 
from several X-class solar flares by satellite-borne or 
ground-based detectors~\citep{Chupp_800621,Chupp_820603,910604_NT,D_900524_1,910604_NM}.
The first clear detection was achieved with the 
Gamma-Ray Spectrometer (GRS) on board the Solar Maximum Mission (SMM), on the occasion of an 
X2.6 solar flare on 1980 June 21~\citep{Chupp_800621}. 
On the other hand, all ground-based positive detections were associated with more energetic solar flares, 
with the GOES class larger than X8. 
While these successful detections indeed demonstrated the production of solar neutrons in solar flares,
it was thought that the neutron detection from medium X-class flares would be difficult with ground-based observations.

Surprisingly, solar neutrons were successfully detected from an X2.3 solar flare on 2000 November 24, 
at near the maximum of the solar cycle 23, 
by a neutron monitor installed at Mt. Chacaltaya in Bolivia~\citep{W_001124}. 
The Bolivian neutron monitor is situated at an altitude of 5250 m and has an area of 13 $\mathrm{m^{2}}$. 
This observation therefore inspired that ground-based detectors at very high mountains
would be able to detect solar neutrons even from medium X-class solar flares, although the 2000 November 24 event might be exceptional. 

Because of the small ($\sim$10) number of positive detections of solar neutrons, 
it is not clear at present how the arrival flux of solar neutrons depends on the flare intensity. 
Furthermore, it is unlikely that the neutron detections will increase in number significantly in near future.
Therefore, we may utilize even upper limits on the solar-neutron flux for a much larger sample of solar flares.
These upper limits at the Earth are expected to constrain some key parameters of the underlying ion 
acceleration process in solar flares, including the total number, the energy content, and pitch angle distributions of the accelerated ions.
In the present paper, we hence analyze the data from the Yangbajing neutron monitor for any signals associated with
166 M- and X-class flares, detected with the BATSE (Burst And Transient Source Experiment) on board Compton Gamma Ray Observatory (CGRO)
over a period of 1998 October through 2000 June.

In Section 2 we give an overview of the Yangbajing neutron monitor, and describe in Section 3 its performance using Monte Carlo simulations.
Section 4 presents our methods used in this analysis. The derived results are given in Section 5, together with 
a brief discussion.
\section{The Yangbajing neutron monitor}\label{sec:YBJNM}
The Yangbajing neutron monitor (NM) has been operated at Yangbajing ($90{\degr}.522$E, $30{\degr}.102$N) 
in Tibet, China,
since 1998 October~\citep{YBJ_NM1999, YBJ_NM2001}.  Its high altitude, 4300 m above sea level, 
provides a much reduced air mass, 606 $\mathrm{gcm^{-2}}$. 
It consists of 28 NM64 type detectors~\citep{NM64_1,NM64_2}, 
attaining a total area of 31.7 $\mathrm{m^{2}}$ which is the largest one among the world-wide NM network at present.
Another advantage of the Yangbajing NM is that it has the highest geomagnetic cutoff rigidity, 14 GV, among NMs
in the world, thanks to geomagnetic conditions. 
These conditions make the Yangbajing NM one of the most sensitive detectors for solar neutrons.

\section{Performance of the Yangbajing neutron monitor}\label{sec:perf}
The data acquisition system of the Yangbajing NM records the event number of
each counter every second. The counting rate is typically 100 Hz per counter, so the total rate
becomes around 2.8 kHz.
In order to examine whether this value is reasonable,
we calculated the expected counting rate of the Yangbajing NM by means of a full Monte Carlo (MC)
simulation in the atmosphere called Cosmos uv6.35~\citep{cosmos}.
Primary cosmic-ray particles were sampled from the energy spectrum constructed from direct observations
in the energy range from 10~GeV to 100 TeV~\citep{JACEE,BESS_1,RUNJOB,BESS_2}. 
An error of $\pm 20$\% was assigned to the absolute flux of primary particles in lower energies, 
whereas its uncertainty was set to be $\pm 50$\% at around 100~TeV. 
These primary particles were thrown isotropically within the zenith angle
$0\degr$ to $60\degr$ on top of the atmosphere, and then fluxes of various secondary particles at 
the altitude of Yangbajing were evaluated.
The flux of species of each secondary particle was finally multiplied 
by the detection efficiency of the NM64 detector calculated by \citet{NM_det},
which in turn was confirmed via comparison with an accelerator experiment using 100 $-$ 400~MeV neutron beams~\citep{RCNP_2}.

As a consequence of the calculation, the expected counting rate of the Yangbajing NM was estimated as 3.4 kHz.
This is in agreement with the experimental counting rate of 2.8 kHz within the error of
the absolute flux of primary particles. 
In addition, the MC simulation revealed how various secondary particle species contribute to the total counting rate.
The secondary neutrons were found to dominate the counting rate with a typical fraction of 83\%. 
The next dominant contribution comes from protons, 15\% to the total counting rate. 
The remaining marginal 2\% is mainly due to secondary negative muons that
are captured in the lead blocks through a reaction of $\mu^{-} + \mathrm{p} \rightarrow \mathrm{n} + \nu_{\mu}$.
This result ensures that 98\% of the counting rate is dominated by secondary nucleons, as expected.

\begin{figure}
  \centering
  \includegraphics[width=10cm]{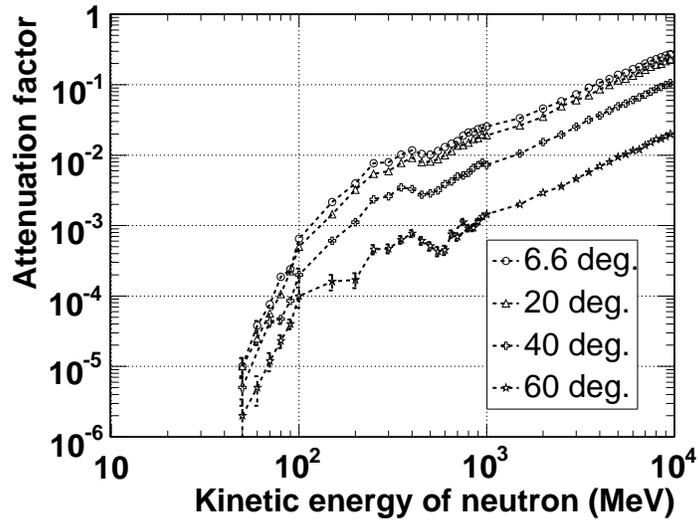}
  \caption{The atmospheric attenuation factor of solar neutrons, calculated via Monte-Carlo simulations 
    assuming different incident zenith angles, and shown 
    as a function of kinetic energy of neutrons measured at the top of the Earth atmosphere. }
  \label{fig:attenuation}
\end{figure}

For a further analysis, we also derived an attenuation factor of solar neutrons in the atmosphere. 
The attenuation of solar neutrons is caused by their collisions with atmospheric nuclei
during their propagation to the observers, depending on their kinetic energies and incident angles.
This is illustrated in Fig.~\ref{fig:attenuation}, which was obtained using  MC
simulations of CORSIKA 6.500~\citep{corsika} embedding GHEISHA~\citep{gheisha,gheishafix} 
as a low-energy ($<$ 80~GeV/n) hadronic interaction model.
Thus, low-energy solar neutrons are strongly reduced in flux, particularly when the incident zenith angle is large.
For example, solar neutrons with a kinetic energy of 100~MeV and an incident angle of $60{\degr}$
suffer heavy attenuation by four orders of magnitude.
Even with such a heavy attenuation, we expect successful detections of solar neutrons with the Yangbajing NM, 
if the arriving neutron flux exceeds $\sim 1 \mathrm{cm^{-2}s^{-1}}$ as found in some past solar flares
with the GOES class above X10, and if each event lasts typically $\sim$ 10 minutes.

\section{Analysis}\label{sec:analysis}
\subsection{Flare sample definition}\label{sec:flare_def}
Since the beginning of its operation on 1991 April 19 through the reentry to the atmosphere on 2000 June 4, 
the BATSE aboard CGRO observed more than 7,000 solar flares in the hard X-ray 
range above $\sim $ 25 keV; the flare list is publicly available\footnote{ftp://umbra.nascom.nasa.gov/pub/batse/events/}. 
Among them, 1205 events were detected over a period of 1998 October and 2000 June, i.e., overlapping 
with the Yangbajing NM operation, with the GOES peak flux ($F_\mathrm{GOES}$) higher than 1.0 $\times 10^{-6}$ $\mathrm{W m^{-2}}$ 
which corresponds to the GOES class of C1.
The flare number reduces to 1013 when we exclude those events of which the Yangbajing NM did not operate due to maintenance
or system troubles.  The 1013 events,  composed of 847 C-class, 159 M-class, and 7 X-class flares, 
constitute our ``very preliminary sample'', which is further divided into two subsets.
One subset is called ``preliminary sample'', consisting of the 166 flares with the GOES class larger than M1. 
The other is ``sub-preliminary sample'',  with the 847 C-class flares.

\begin{figure}
  \centering
  \includegraphics[width=10cm]{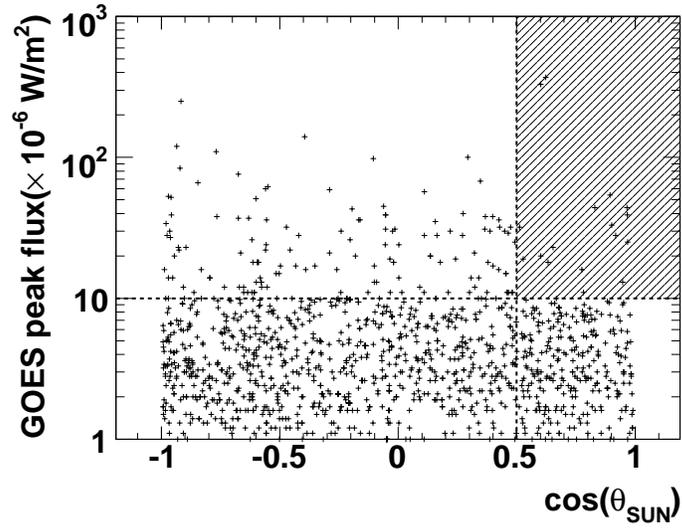}
  \caption{
    The cosine of the zenith angle of the Sun ($\theta_\mathrm{SUN}$) at Yangbajing
    at the onset time of each flare in our very preliminary sample, 
    plotted against their GOES peak flux in units of $10^{-6}$ $\mathrm{Wm^{-2}}$. 
    The vertical dashed line indicates $\theta_\mathrm{SUN} = 60{\degr}$, while the horizontal
    one represents the GOES flux of 1 $\times 10^{-5}$ $\mathrm{Wm^{-2}}$, or the GOES class of M1. 
    The 18 events in the hatched area constitute our final sample.
  }
  \label{fig:zenith_class}
\end{figure}

As indicated by Fig.~\ref{fig:attenuation}, the zenith angle of the Sun at Yangbajing, $\theta_{\mathrm{SUN}}$, 
must be small in order for the Yangbajing NM to detect neutron signals. Therefore, we arranged the 1013 ``very preliminary''
sample events, in Fig.~\ref{fig:zenith_class}, on the plane of $\cos (\theta_\mathrm{SUN})$ calculated at the flare onset vs $F_\mathrm{GOES}$.
Based on this plot, we have selected 18 flares for 
our ``final sample'', with criteria that $\theta_{\mathrm{SUN}}$ is smaller than $60{\degr}$ and $F_\mathrm{GOES}$ is higher than 1.0 $\times 10^{-5}$ $\mathrm{Wm^{-2}}$. 
These flare are summarized in Table~\ref{tab:list_solar_flares}.

\subsection{NM light curves}\label{sec:BG}
\begin{figure*}
  \centering
  \includegraphics[width=10cm]{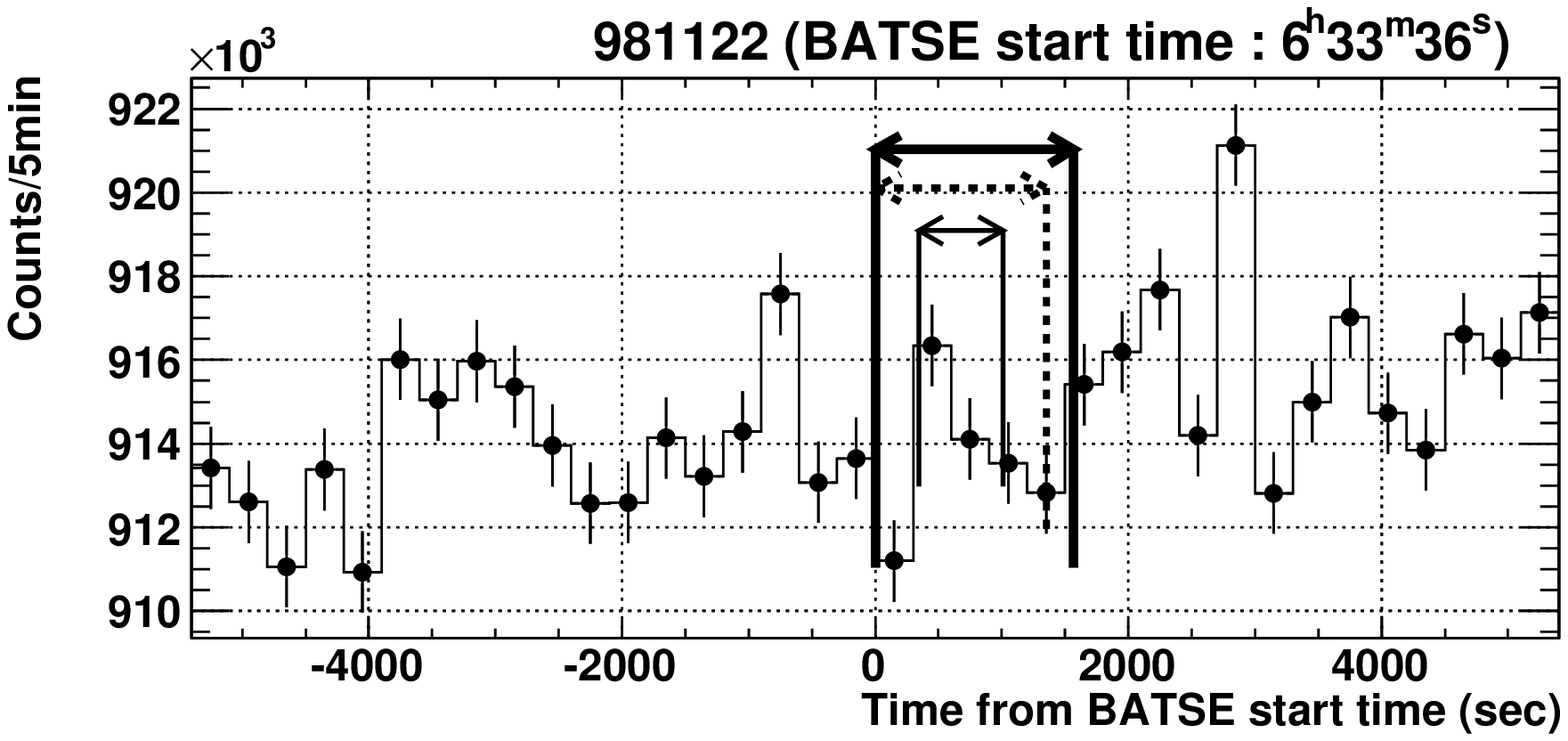}  \includegraphics[width=10cm]{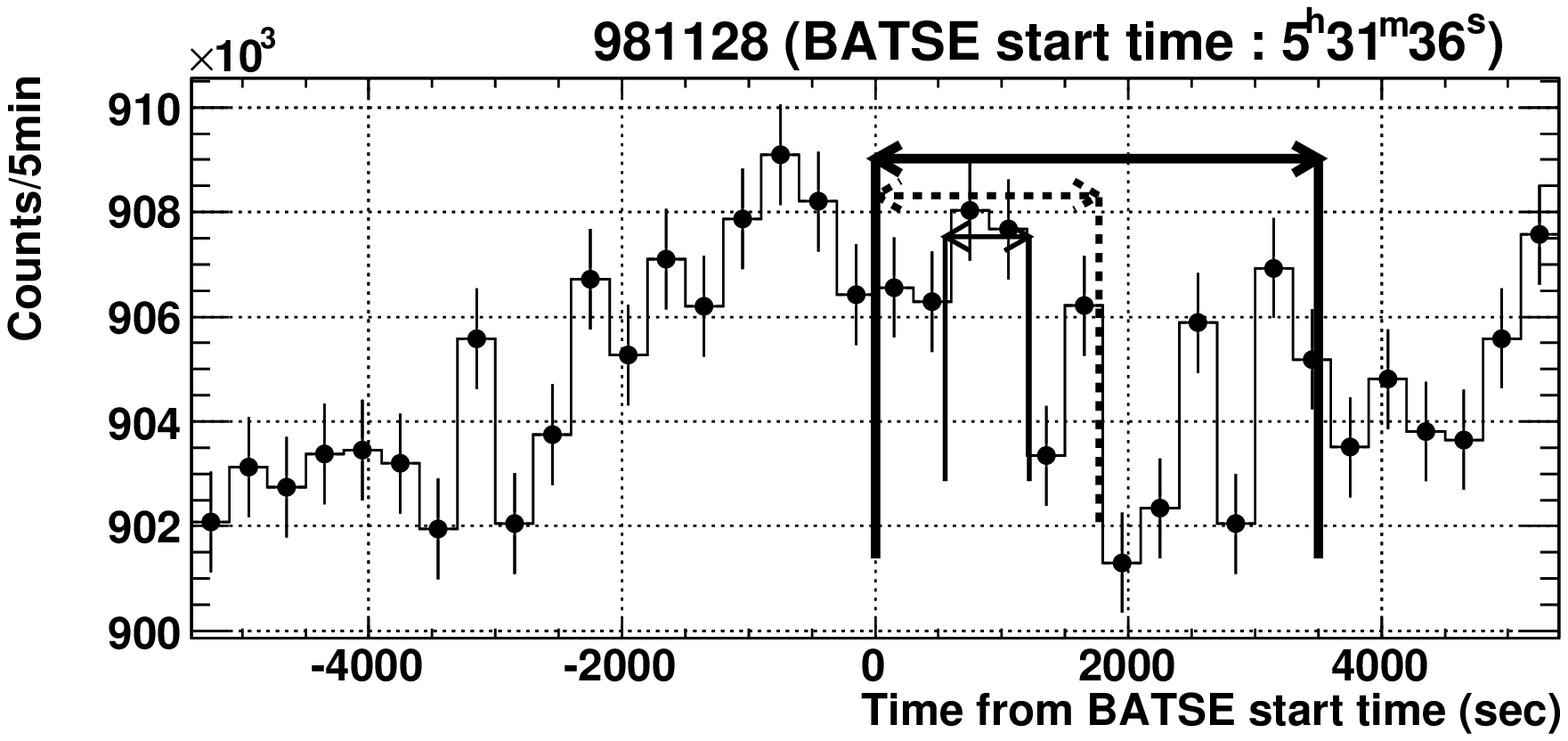}
  \includegraphics[width=10cm]{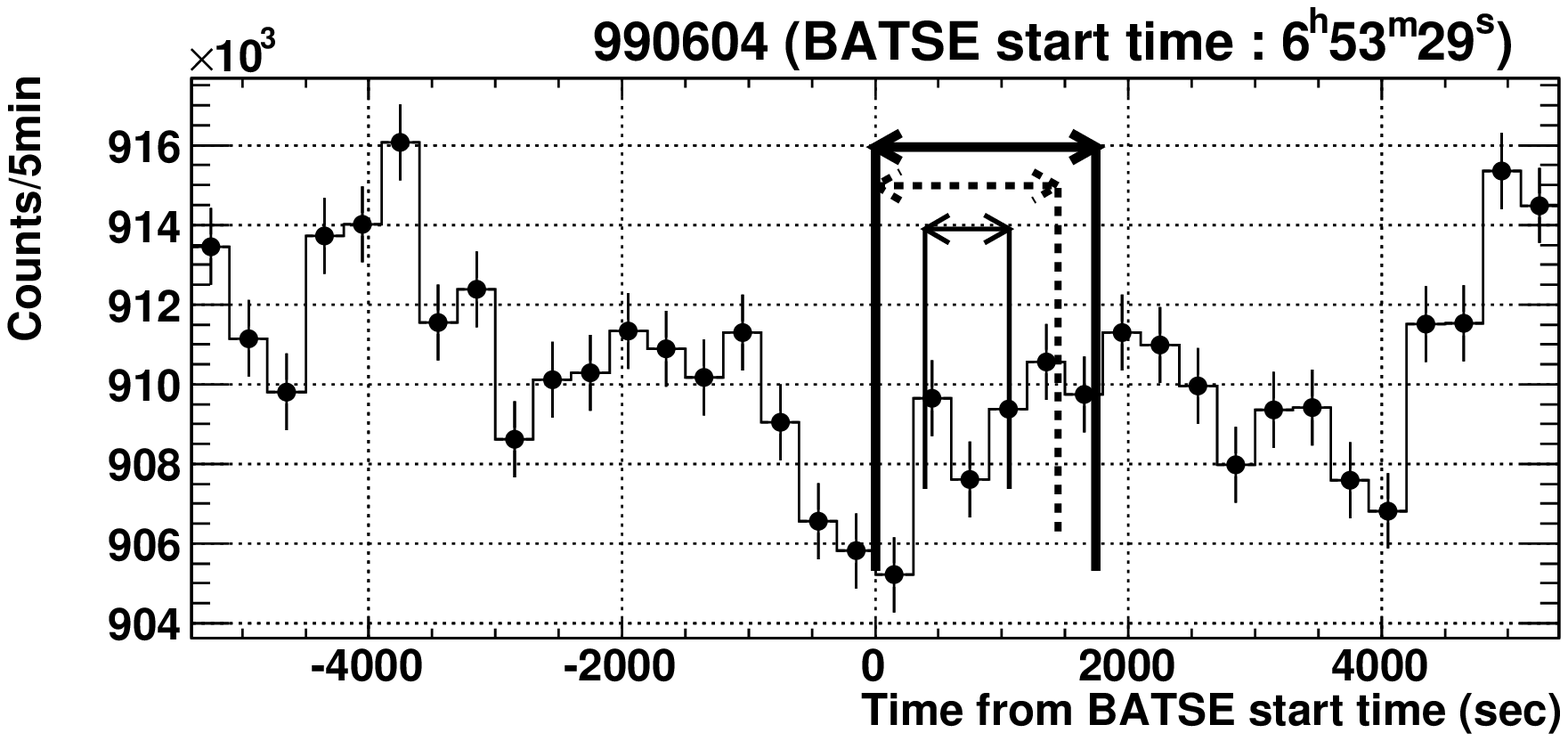}  \includegraphics[width=10cm]{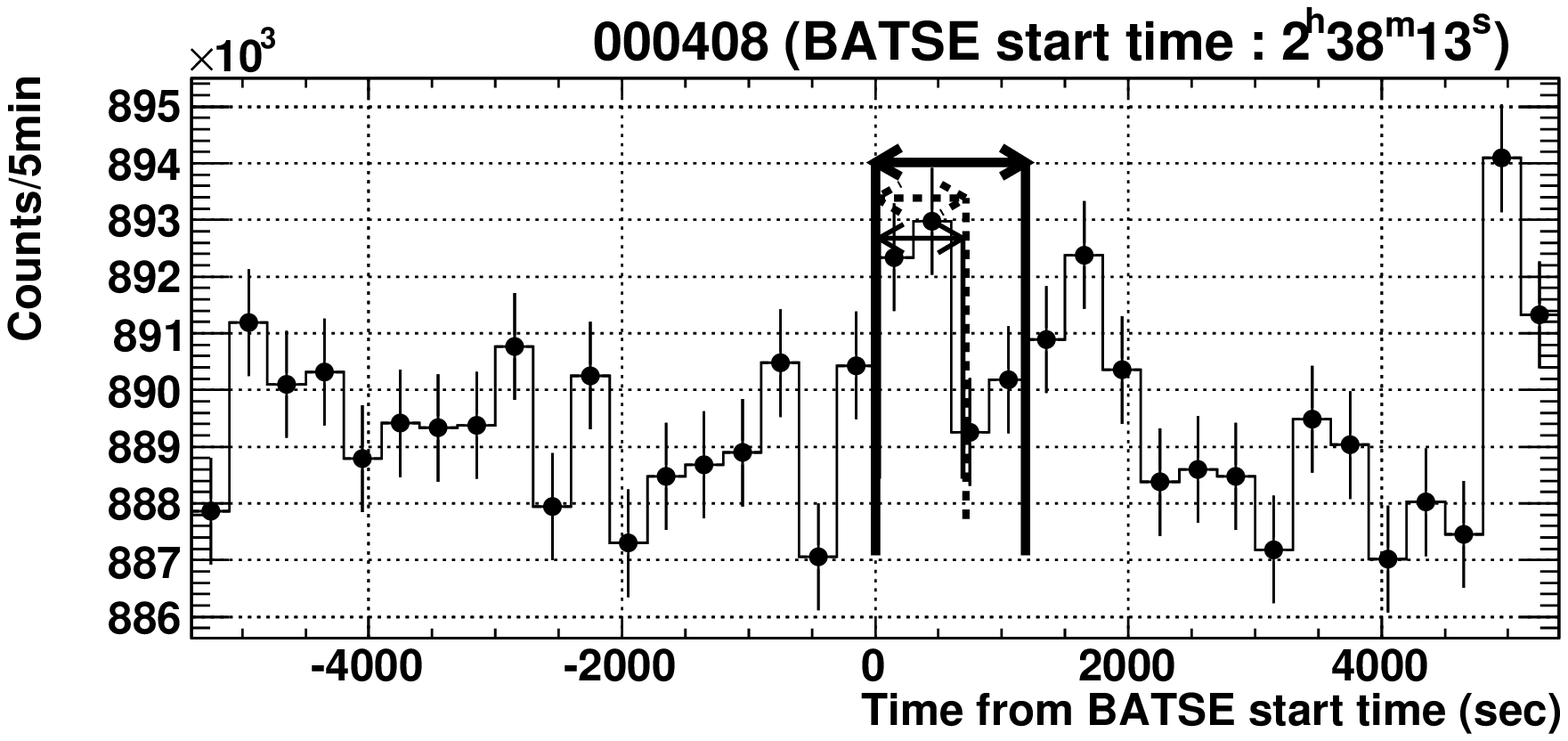}
  \caption{
    The 5-minute counting-rate histories of the Yangbajing NM for two X-class and two M-class flares.
    Top and second panels show light curves for an X3.7 (981122) and X3.3 (981128) flares, respectively, 
    while third and bottom panels indicate those for an M3.9 (990604) and M2.0 (000408) flares, respectively.
    Abscissa in all panels shows the time measured from $t_\mathrm{S}$, ranging over $\pm$ 5400s, where zero corresponds to 
    the BATSE start time (Table~\ref{tab:list_solar_flares}).
    The intervals of individual ON time windows specified by the $\delta$-, continuous-, and gaussian-emission profiles are indicated 
    by horizontal thin, thick, and dashed arrows, respectively.
    The errors indicate uncorrected $1\sigma$ statistical ones, simply calculated by 
    square root of each 5-minutes counts.
} 

  \label{fig:sample_def}
\end{figure*}

Our basic strategy is to search the NM count-rate histories (or ``light curves'') for 
possible excess counts associated with solar flares. For each flare in our preliminary sample,
we therefore prepared a light curve from the Yangbajing NM, spanning $\pm$ 1.5 hours from the flare onset
time, $t_\mathrm{S}$, determined by the BATSE data. This time interval is considered long enough to cover each flare, and
to estimate background count levels. We binned the light curves into 5 minutes per bin, and corrected them
for atmospheric pressure changes which cause variations in the attenuation factor by typically $\pm$2\%. 
Figure~\ref{fig:sample_def} shows four particular examples of such light curves out of the 18 final-sample flares;
these are two X-class flares, 981122 and 981128, 
and two M-class ones, 990604 and 000408,  with the latter two being the highest and the second 
highest BATSE peak rates among M-class flares in 
our final sample (Table~\ref{tab:list_solar_flares}). Thus, no excess counts due to solar neutrons are apparently observed.

In order to more quantitatively constrain neutron counts associated with each flare, we need to define for each flare
an ``ON time window'', i.e., a time interval when solar neutrons might indeed arrive at the Yangbajing NM, 
and use the remaining two time intervals (before and after the flare) to estimate the background. For this purpose,
we must in turn assume a time profile of the solar-neutron production at the Sun in each flare,
as well as the maximum and minimum kinetic energies of the produced
solar neutrons. Then the ``ON time window''opens at the arrival of the most energetic neutrons 
ejected at the beginning of the neutron emission at the Sun, and closes at that of the least energetic ones ejected
at the end of the production interval. 
In this work, we postulate that the maximum and minimum kinetic energies of 
solar neutrons are 10~GeV and 100~MeV, respectively.
				   
The following three neutron-emission time profiles are employed;
(1) $\delta$-emission, (2) continuous-emission, and (3) gaussian-emission. 
the $\delta$-emission simply means that solar neutrons are emitted from the Sun 
instantaneously at the BATSE HXR emission peak, 
while the continuous-emission profile assumes that neutrons are continuously and constantly 
radiated from the Sun throughout the BATSE HXR emission. 
Here for the purpose of introducing mathematical forms of each emission profile, we define ``re-normalizing'' time $t$ as $T - t_\mathrm{S}$,
where $T$ and $t_{\mathrm{S}}$ show the normal time measured in UT and the onset time of the BATSE emission, respectively.
With the renormalizing time $t$ and above assumptions, the neutron emission profile $\nu(t)$ takes either of the following forms;
$\nu(t) = 1$ only at $t = t_{\mathrm{P}} - t_{\mathrm{S}}$ and otherwise 0 if the $\delta$-emission is adopted,
$\nu(t) = 1$ for $t=$ 0 to $\Delta t_\mathrm{b}$ and otherwise 0 if the continuous-emission profile is used.
Here, $t_\mathrm{P}$ is the peak time of the BATSE emission, and $\Delta t_\mathrm{b}$ is the BATSE flare duration.
Hereafter we define $T_\mathrm{PS}= t_{\mathrm{P}} - t_{\mathrm{S}}$. 
Both these cases have been considered in some past studies~\citep{Chupp_800621,910604_NT,Shibata1994,W_001124,B_031028}. 

In comparison with these simplified emission profiles, 
the gaussian-emission profile is a rather realistic one, defined as 
\begin{equation}
    \nu(t) = 
    \left\{
    \begin{array}{ll}
      C \exp \left( {-(t - T_\mathrm{PS})^{2}/2\Delta T^{2} }\right)  & \mbox{if $0 \le t \le 2T_\mathrm{PS}$},   \\
      0                                                        & \mbox{otherwise},
    \end{array}
    \right.,
    \label{eq:gauss-emission}
\end{equation}
where $C$ is a normalization factor so that time integral of $\nu(t)$ 
over 0  -- $2T_\mathrm{PS}$ is unity, and its centroid and standard deviation $\Delta T$
are set to be $T_\mathrm{PS}$ and $0.5T_\mathrm{PS}$, respectively.
This is because that light curves of the 4 -- 7~MeV nuclear gamma-ray emission, 
which obviously reflect ion acceleration at the Sun, is approximately described 
by a Gaussian function~\citep[e.g.,][]{Chupp_cycle21,W_001124}.
Furthermore, according to various solar X-ray and gamma-ray observations~\citep[e.g.,][]{FC1983_800621,W_031028}, 
the rise and peak times of the 4 -- 7~MeV nuclear gamma-ray light curve are roughly 
similar to those of the HXR emission. 

In this work we treat three neutron emission profiles, 
since it is natural that ions are accelerated in solar flares at the same time as electrons. However, 
a recent work performed by~\citet{sako_20050907} suggests that acceleration time or trapping time 
of ions in 2005 September 7 flare may be longer than that of electrons. Thus, it should be noted that 
we might expect more arrival neutron flux than the following investigation.

The ON time windows specified by these three profiles are illustrated in Fig.~\ref{fig:sample_def}.
Here we give a brief explanation of the ON time windows, using 981122 flare 
(top panel of Fig.~\ref{fig:sample_def}) as an example.
Since the $\delta$-emission supposes instantaneous neutron emission at the Sun,
the corresponding ON time window starts at $t_\mathrm{P}+  1.85$ s and end at $t_\mathrm{P} + 667$ s, lasting for 665 s, 
where 1.85 s and 667 s are time delays of 10~GeV and 100~MeV neutrons, respectively, with respect to photons.
Since this flare has $T_\mathrm{PS} = 341$ s, the ON time windows lasts from 343 s to 1008 s as measured from $t_\mathrm{S}$.
The ON time windows specified by the other two models are longer than that by the $\delta$-emission, 
because these assume prolonged neutron emissions. It should be stated that 5-minute counts shown in Fig.~\ref{fig:sample_def}
seem to be highly variable. This is because normal Poisson error is assigned to each count. In order to 
explain that the variations are not so large, we introduce a correction factor for errors of 5-minute counts 
in Section \ref{sec:cal_sig}.

By excluding all count bins in the ON time window and two adjacent count bins, 
two time regions are available in each light curve for background estimation. 
We then determined the background level of the NM counts, by fitting a common quadratic curve to 
these two time windows through a least square method. 
The calculated quadratic curve in turn enables us to define the background 
level inside the ON time window. This removes 
residual temporal trend (such as seen in the 3-hours data of Fig.~\ref{fig:sample_def}), 
mostly due to solar diurnal variation with a typical amplitude of $\pm$ 0.3\%;
this arises from diffusion and convection of Galactic cosmic-rays below several tens GeV in the interplanetary space~\citep[e.g.,][]{Parker_1965,JP_1970}.
Finally, all counts in the ON time window are summed up (hereafter $N_\mathrm{ON}$), and 
all the interpolated background counts in the same ON time window are likewise accumulated (hereafter $N_\mathrm{BG}$).
Then the signal from solar neutrons in the ON time window, $N_\mathrm{S}$, is available by 
simply subtracting $N_\mathrm{BG}$ from $N_\mathrm{ON}$. 
Here the $N_\mathrm{S}$ would probably correspond to the ``worst'' neutron signals since
we equally treat all counts in an ON time window as neutron signals. Practically, more neutron signals could be expected 
via optimizing an ON time window for any emission models. This is our future work.
\subsection{Significance of the neutron signal}\label{sec:cal_sig}
For the purpose of calculating the statistical significance of $N_\mathrm{S}$, we need to estimate
its uncertainty $\Delta N_\mathrm{S}$ given by $\sqrt{\Delta N_\mathrm{ON}{}^{2} + \Delta N_\mathrm{BG}{}^{2}}$,
where $\Delta N_\mathrm{ON}$ and $\Delta N_\mathrm{BG}$ represent errors associated with 
$N_\mathrm{ON}$ and $N_\mathrm{BG}$, respectively.
However these quantities do not obey Poisson statistics,
since the lead blocks, used to increase the detection efficiency through multiplication of incident neutrons, 
often lead to multiple counts in the $\mathrm{BF_{3}}$ counter for one incident neutron.  
This prevents us from estimating, e.g., simply as $\Delta N_\mathrm{ON} = \sqrt{N_\mathrm{ON}}$.
Therefore, using the NM data themselves, let us derive a correction factor $\Gamma(> 1)$ 
which yields $\Delta N_\mathrm{ON} = \Gamma \sqrt{N_\mathrm{ON}}$. 

We first constructed occurrence histograms of the 5-minutes counts each day,
excluding those data which were obtained during 3 days after the occurrence of a preliminary-sample flare.
This is to avoid possible flare-associated effects, such as the arrival of solar energetic charged particles 
and/or a subsequent Forbush decrease. The produced histograms are refereed to collectively as ``background-histogram sample''. 
Then each histogram in the background-histogram sample is fitted with a Gaussian function, in order
to derive its mean value, $\mu$, and standard deviation, $\sigma$, and to consequently obtain the factor 
$\Gamma = \sigma/\sqrt{\mu}$. Here such a multiplicity in NMs was firstly discussed by~\cite{NM64_1}.

\begin{figure}
  \centering
  \includegraphics[width=8cm]{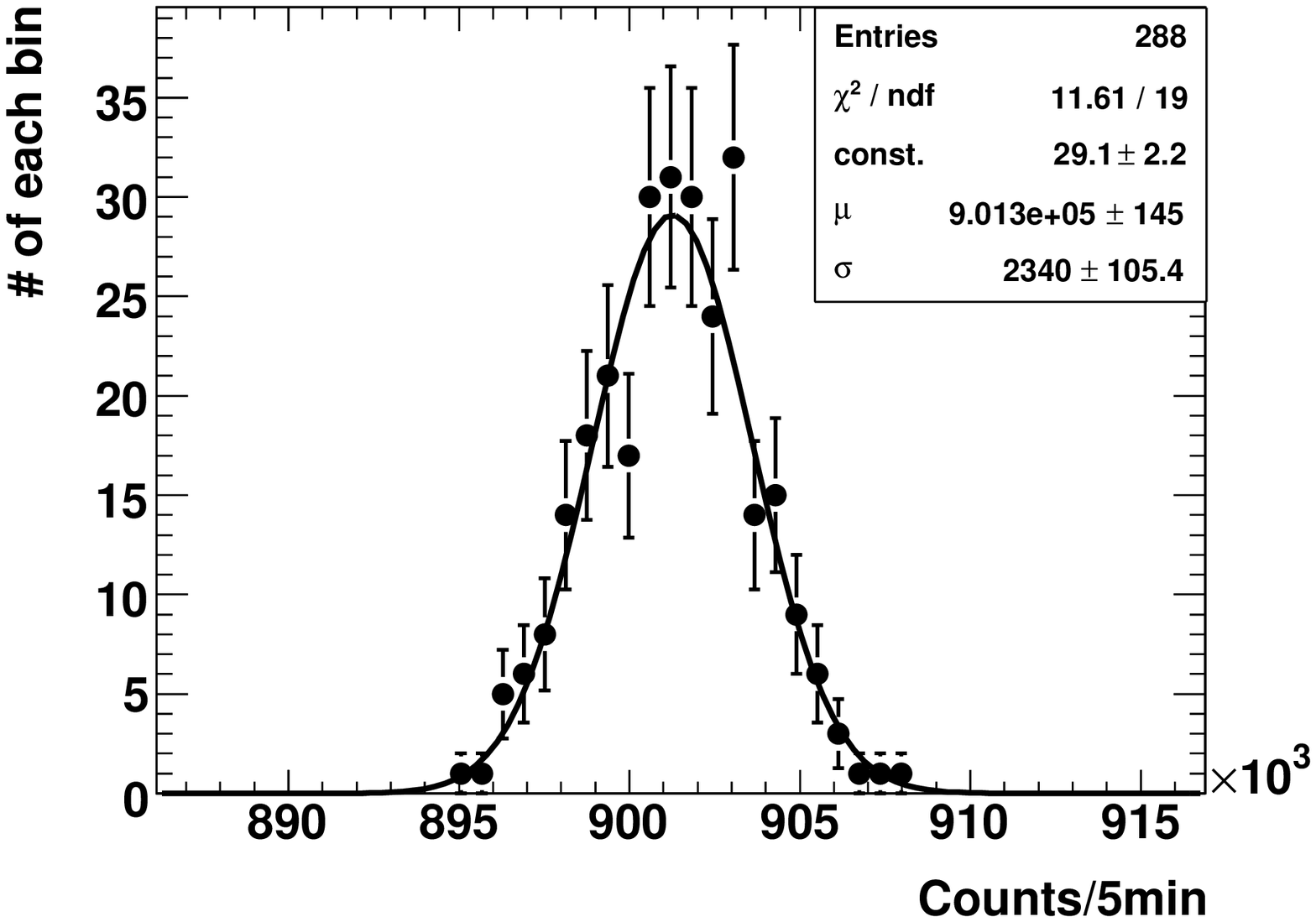}
  \includegraphics[width=8cm]{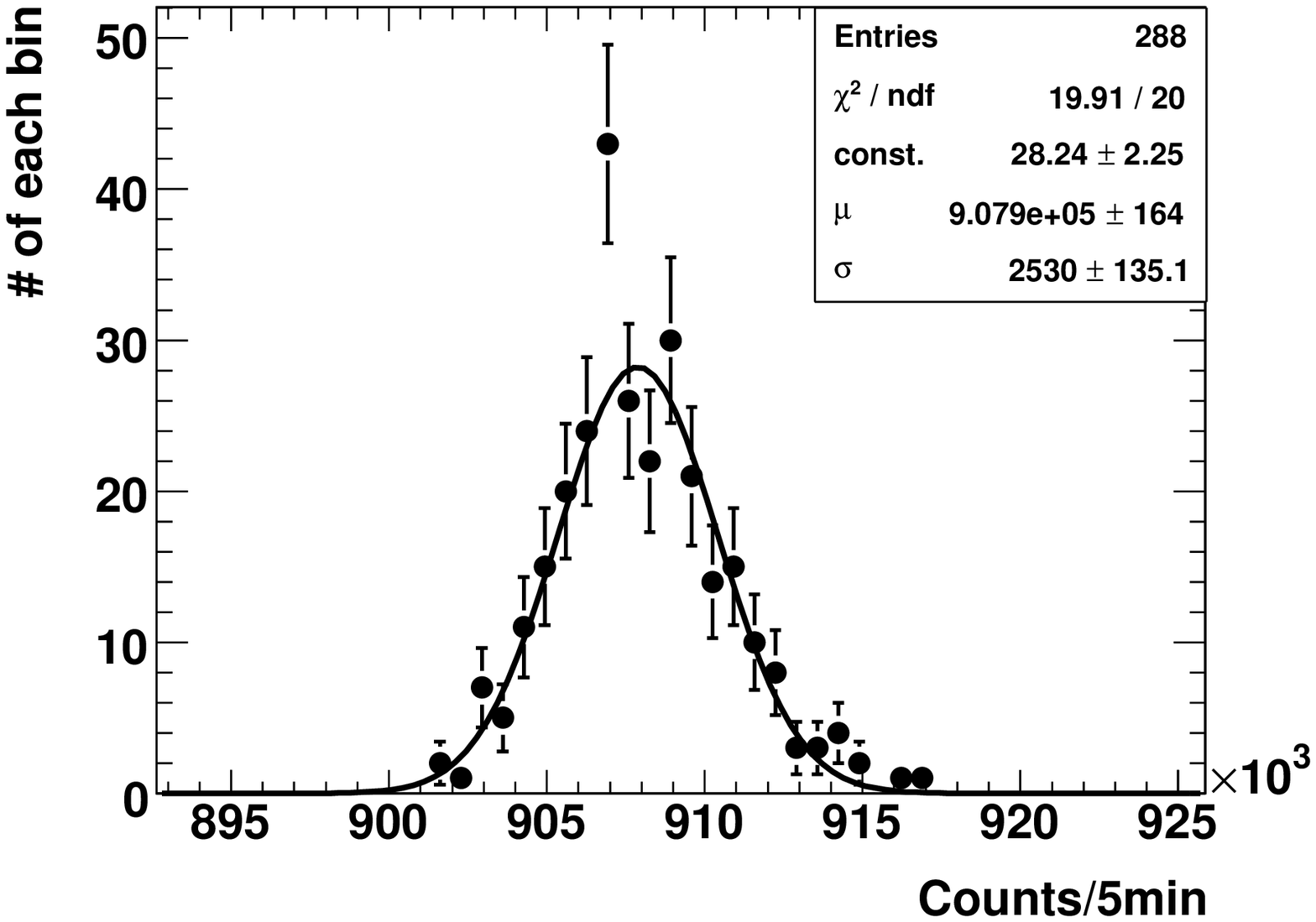}
  \includegraphics[width=8cm]{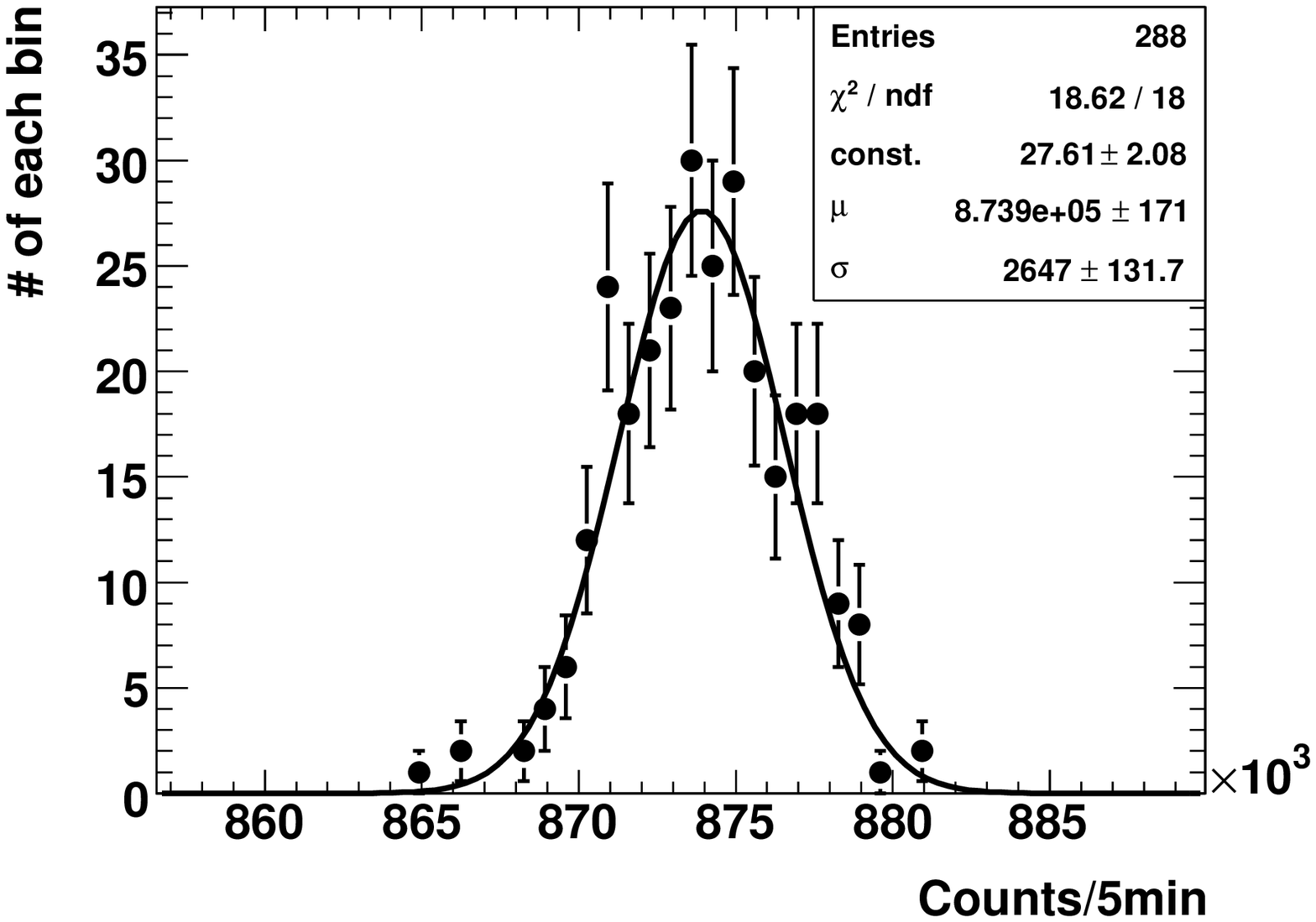}
  \caption{Typical occurrence histograms of the 5-minutes counts 
    of the Yangbajing NM data on 1998 December 3 (top),
    1999 June 3 (middle), and 2000 April 30 (bottom).
    Error bars assigned to individual data points are Poissonian $\pm 1\sigma$. 
    The solid line indicates the best Gaussian fit to each distribution. 
  }
  \label{fig:typical_countdist}
\end{figure}

Figure~\ref{fig:typical_countdist} shows some typical distributions in the background-histogram sample. 
Since the distribution of top panel of Fig.~\ref{fig:typical_countdist} has $\mu = 901270 \pm 145$ 
and $\sigma = 2340 \pm 105$, both per 5 minutes,
we find that $\sigma$ deviates from the expected Poisson fluctuation, $\sqrt{\mathrm{\mu}}$, 
by a factor of $\Gamma = 2340/\sqrt{901270} = 2.46 \pm 0.11$. 
The other distributions shown in Fig.~\ref{fig:typical_countdist}
also exhibit similar excess above the expected Poisson fluctuation, yielding $\Gamma = 2.65 \pm 0.14$ 
(middle panel) and $2.83 \pm 0.14$ (bottom panel).
Deriving the factor $\Gamma$ from each histogram in the background-histogram sample in the same way,
and averaging them over the analyzed period, we obtained the averaged correction factor as $\Gamma = 2.73 \pm 0.01$. 
Furthermore, the standard deviation in 5-minutes counts has been estimated as $\sigma^{*} = 2576 \pm 16$.
As can be seen from the particular examples of Fig.~\ref{fig:typical_countdist}, 
the mean $\mu$ varies by $\pm 10 - 20$\% from day to day,
due, e.g., to monthly and yearly solar activity changes. The factor $\Gamma$ also 
varies by a similar extent, depending on the day.
However we regard $\Gamma$ as constant, because the fluctuations in $\Gamma$ do not affect our final results.
Here, as a general remark, a recent study for NM multiplicity given by ~\citet{B_NMmul} has 
suggested that the multiplicity for a 100 MeV neutron are relatively larger than
the derived value in this work. Probably, the difference may result from some systematic differences such as
logic of electronics, arrangement of NMs, and geophysical conditions of a station.

Using $\Gamma$ as calculated above, $\Delta N_\mathrm{ON}{}^{2}$ is determined as 
$\sum (\Gamma \sqrt{N_\mathrm{cnt}})^{2} = \Gamma^{2} N_\mathrm{ON}$, 
where summation is done over an ON time window,
while $\Delta N_\mathrm{BG}{}^{2}$ is evaluated as $\sum (\sigma^{*}/\sqrt{n_\mathrm{off}})^{2}$, where
summation is also performed over the ON time window and $n_\mathrm{off}$ is the number of 5-minutes bins 
used to define the background level through the least square method.
Finally, the significance of possible neutron signals, $\sigma_\mathrm{S}$, can be calculated as 
\begin{equation}
  \sigma_{\mathrm{S}} = 
    \frac{N_\mathrm{ON} - N_\mathrm{BG}}{\sqrt{ \Gamma^{2} N_\mathrm{ON} + \sum (\sigma^{*}/\sqrt{n_\mathrm{off}})^{2} }}.
  \label{eq:stat_2}
\end{equation}

\section{Results and discussions}
\subsection{Significance distributions}
With equation~(\ref{eq:stat_2}), we calculated the statistical significances of the 18 final-sample flares. 
Table~\ref{tab:Ns_count} summarizes the values of $\sigma_\mathrm{S}$, for the three neutron production 
profiles, together with $N_\mathrm{S}$ and $\Delta N_\mathrm{S}$. 
Thus, the signal significance is at most $2 \sigma$, implying all null detections. 
Then, how about the other flares ?

\begin{figure*}
  \includegraphics[width=8.5cm]{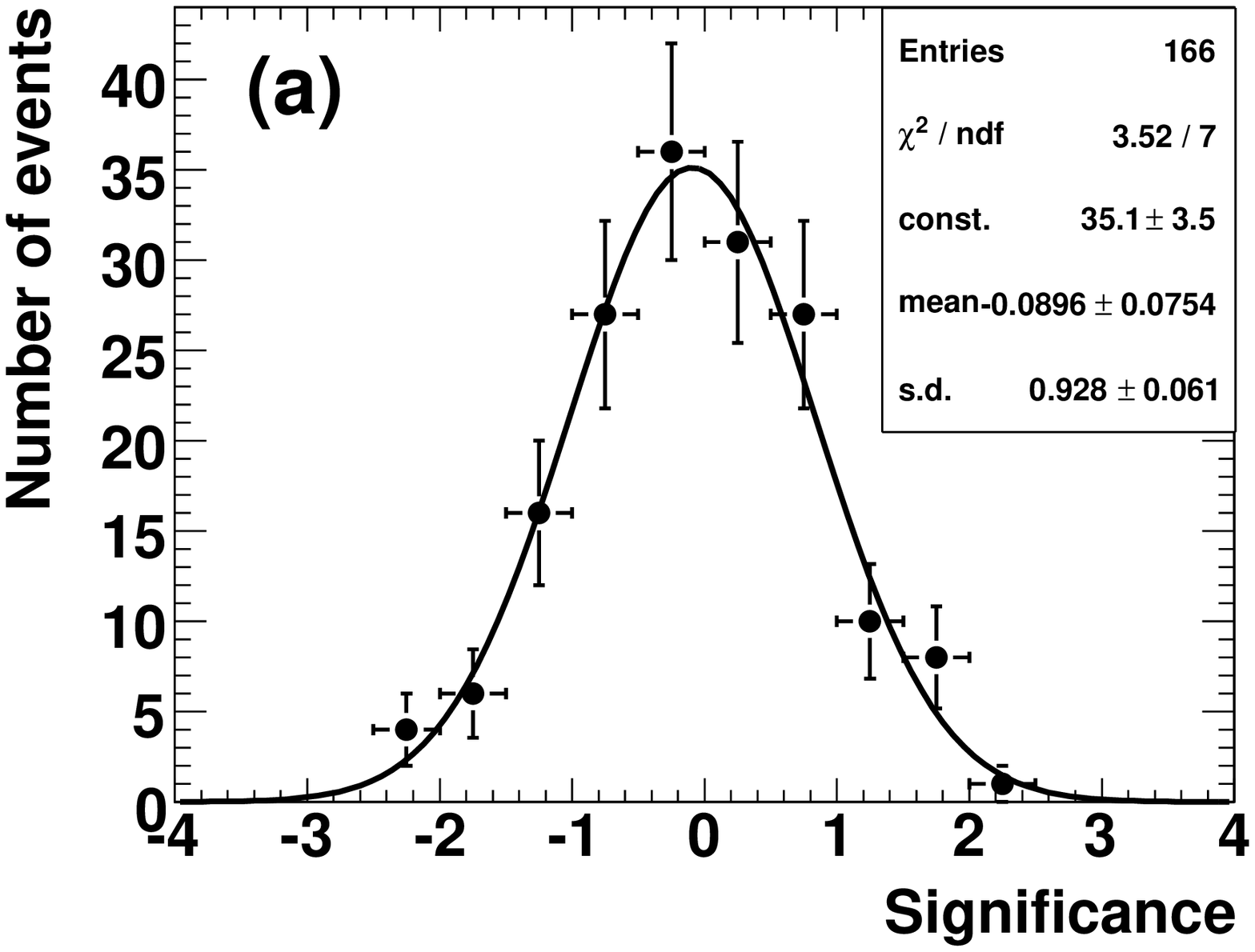}
  \includegraphics[width=8.5cm]{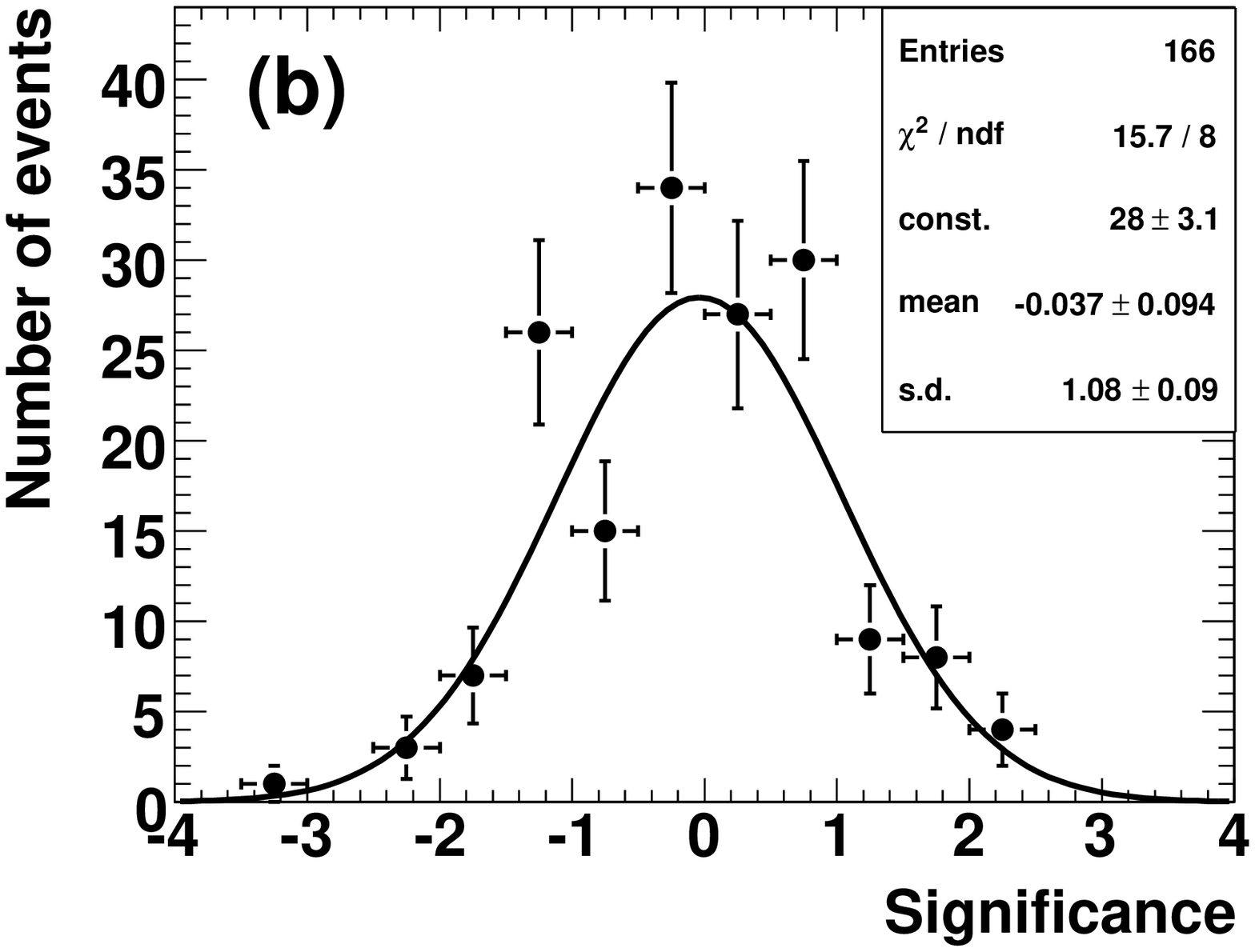}
  \includegraphics[width=8.5cm]{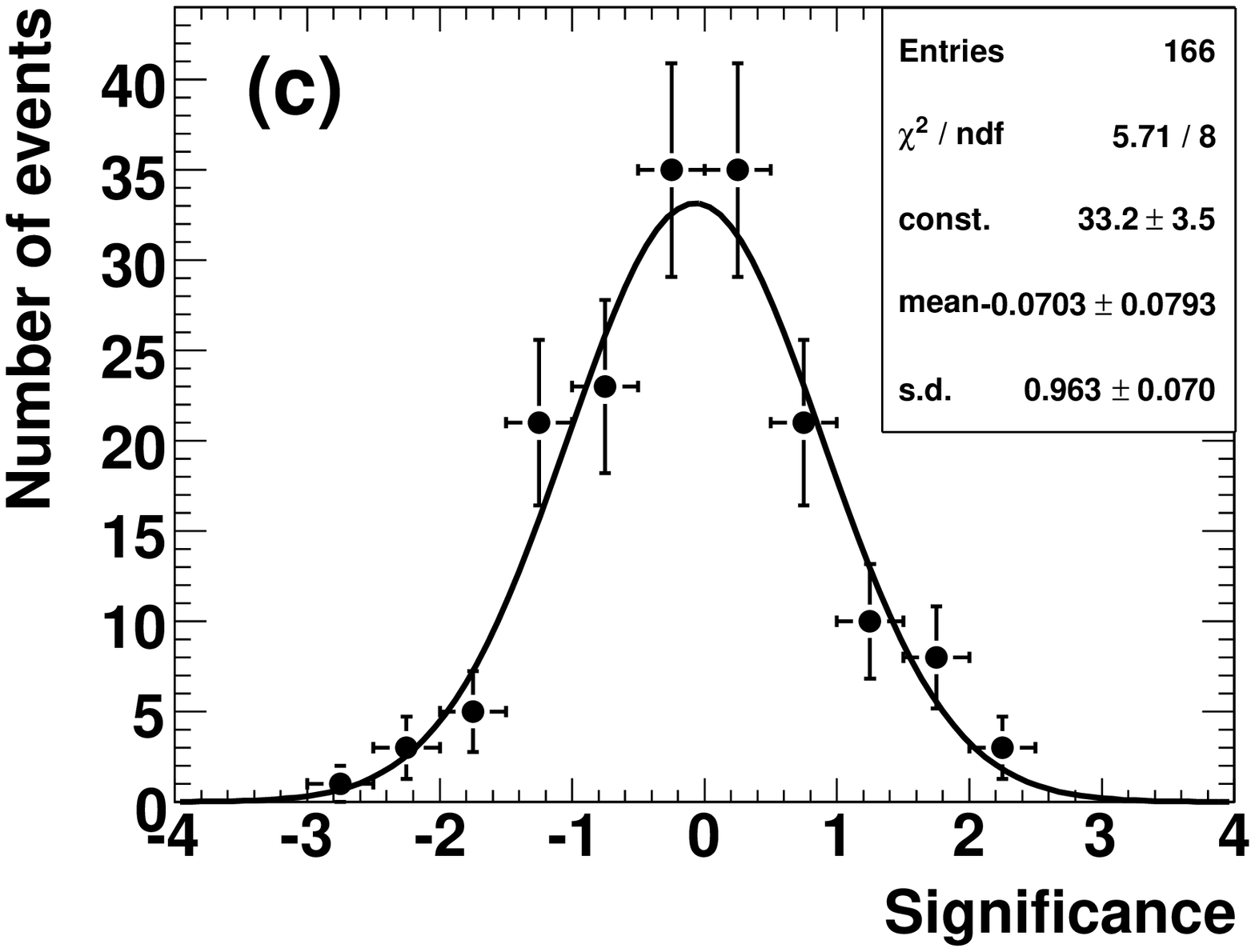}
  \caption{Significance distributions of neutron signals from our preliminary-sample flares, computed assuming the three neutron-emission
    profiles. Panels (a), (b), and (c) correspond  to the $\delta$-, continuous-, and gaussian-emission models, respectively.
    Vertical error bars associated with individual data points are Poissonian.
    The best-fit gaussian curve to each histogram is drawn by a solid line. 
    Statistical parameters obtained from each fit are indicated in the legend of each graph.}
    \label{fig:sigdist}
\end{figure*}

Figure~\ref{fig:sigdist} shows occurrence histograms of the statistical significance of signal neutron counts of our preliminary-sample flares,
calculated with equation~(\ref{eq:stat_2}) in the same way.
The 166 flares have been binned into appropriate intervals.
We fitted each histogram in Fig.~\ref{fig:sigdist} with a Gaussian distribution, with its height, mean value, and standard deviation
left free. The fit has been successful as shown in Fig.~\ref{fig:sigdist}, with two important consequences.
One is that the obtained Gaussian centroids are consistent with 0 within the fitting errors. Therefore, there is
no evidence, in the statistical sense, for positive neutron signals associated with the 166 solar flares. The other point
is that the obtained standard deviations are consistent with 1.0, implying that the significance scatter among 
the 166 flares can be fully explained by statistical fluctuations and that the value of
$\Gamma$ has been correctly estimated. 
The latter fact also ensures that the developed method 
is not affected significantly by any unknown systematic effects. 
For clarifying this, other analysis were performed, 
systematically including one bin before the beginning of the BATSE start time or leaving out the last bin in ON time windows. The difference
between the obtained results shown in Fig.~\ref{fig:sigdist} and those other results can be well explained by statistical fluctuation.
\subsection{Calculation of an upper limit of the solar-neutron flux}\label{sec:cal_ul}
Given the null detections of solar neutrons, 
we proceed to set upper limits on the solar-neutron flux from the 18 final-sample 
flares (Table~\ref{tab:list_solar_flares}).
For this purpose, an upper limit on the neutron counts at 95\% confidence level (CL), 
$N_{95}$, was estimated for each final-sample flare, from the ON time window counts $N_\mathrm{S}$ 
and its uncertainty $\Delta N_\mathrm{S}$, and using a statistical method by \citet{helene}. 
The obtained results on $N_{95}$ are provided in Table~\ref{tab:Ns_count}.

The quantity $N_{95}$ may be written as 
\begin{equation}
  N_{95} = R^{-2} \int^{\Delta t_\mathrm{b} + T_\mathrm{max}}_{0} \!\!\! dt \int^{T_\mathrm{max}}_{T_\mathrm{min}} \!\!\! \nu(t-t')N(E_\mathrm{n})P(E_\mathrm{n})
    A(\theta, E_\mathrm{n})\frac{dE_\mathrm{n}}{dt'}  dt'.
  \label{eq:sn_count}
\end{equation}
Here, $R$ is the distance from the Sun to the Earth; 
$t$ is the time measured from $t_\mathrm{S}$;
$\Delta t_\mathrm{b}$ is again the BATSE duration;
the times $T_\mathrm{min}$ and $T_\mathrm{max}$ correspond to 
delays of 10~GeV and 100~MeV solar neutrons, respectively; 
the variable $t'$ is the difference between the Sun-Earth transit times for neutrons and photons;
$E_\mathrm{n}$ is the kinetic energy of neutrons which is a function of $t'$; and 
$\theta$ is the incident angle of neutrons at the top of the atmosphere, which 
is $\theta_\mathrm{SUN}$ in Table~\ref{tab:list_solar_flares}.
Furthermore, $\nu$ is the neutron production time profile defined in Section~\ref{sec:BG}, 
$N(E_\mathrm{n})$ is the differential neutron spectrum at the Sun, 
$P(E_\mathrm{n})$ is the probability for a neutron to reach the Earth before its decay, and
$A(\theta, E_\mathrm{n})$ is the effective area of the Yangbajing NM which is a product of the
atmospheric attenuation factor (Fig.~\ref{fig:attenuation}) and the detection efficiency 
of an NM64 detector~\citep{NM_det}.
The term $dE_\mathrm{n}/dt'$ is the neutron energy-time dispersion relation~\citep{LR1967}.
We assume the function $N(E_\mathrm{n})$ to have 
a form of $\alpha (E_\mathrm{n}/100$ $\mathrm{MeV})^{-\beta}$ $(\mathrm{sr^{-1}MeV^{-1}})$,
which is theoretically expected from the shock acceleration process of ions in solar flares~\citep[see][ and references therein]{hua2002}.
Here, the power-law index $\beta$ is assumed to take values of 3, 4, and 5 
as conservative ones~\citep{Chupp_820603,910604_NT,910604_NM,W_001124,W_031028}. 
Since $\theta$ varies from flare to flare (Table~\ref{tab:list_solar_flares}), 
$A(\theta, E_\mathrm{n})$ is interpolated from effective areas calculated  for a set of incident angles 
of $6.6{\degr}$, $10{\degr}$, $20{\degr}$, $30{\degr}$, $40{\degr}$,$50{\degr}$, and $60{\degr}$.

On the right hand side of equation~(\ref{eq:sn_count}), the only unknown quantity 
is the spectrum normalization $\alpha$. Therefore, by substituting the measured values 
of $N_{95}$ (Table~\ref{tab:Ns_count}) for the left hand side of equation~(\ref{eq:sn_count}), we can 
determine $\alpha$ for each flare. We can then
calculate a 95\% CL upper limit on the neutron flux at the top of the Earth atmosphere as
\begin{equation}
  F_{95} = \alpha \Delta T_\mathrm{n}^{-1}R^{-2} \int^{\Delta t_\mathrm{b} + T_\mathrm{max}}_{0} \!\!\! dt 
  \int^{T_\mathrm{max}}_{T_\mathrm{min}} \!\!\! \nu(t-t')(E_\mathrm{n}/100 \mathrm{MeV})^{-\beta} P(E_\mathrm{n}) \frac{dE_\mathrm{n}}{dt'}  dt',
  \label{eq:ul_cal}
\end{equation}
where $\Delta T_\mathrm{n}$ is the time interval of the ON time window. 
The inner integral of the right hand side of equation~(\ref{eq:ul_cal}) differs from that of equation~(\ref{eq:sn_count})
in that the effective area $A(\theta, E_\mathrm{n})$ has been eliminated.
\subsection{The flux upper limits of the final-sample flares}\label{sec:all_UL}
\begin{figure*}
  \includegraphics[width=8.5cm]{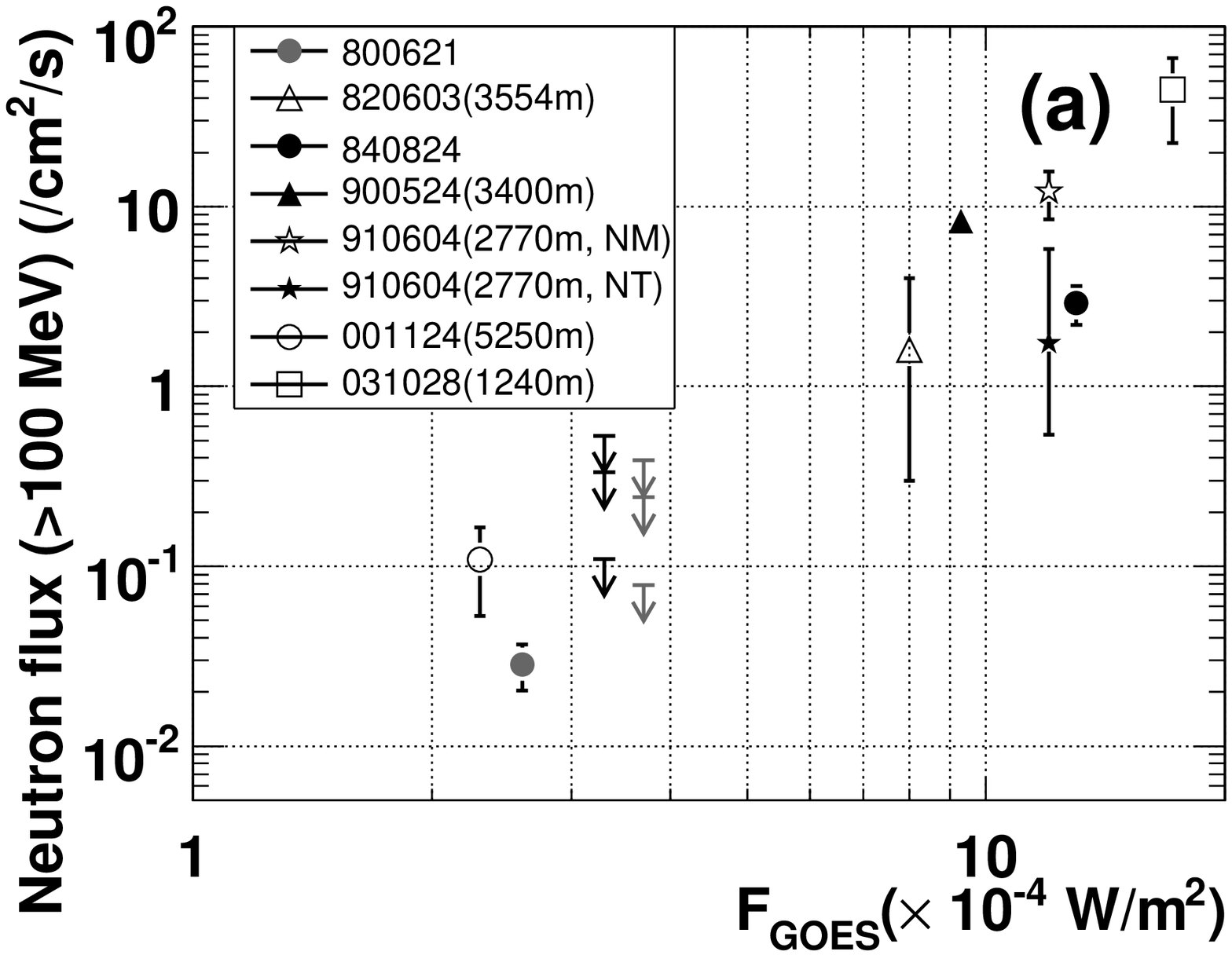}
  \includegraphics[width=8.5cm]{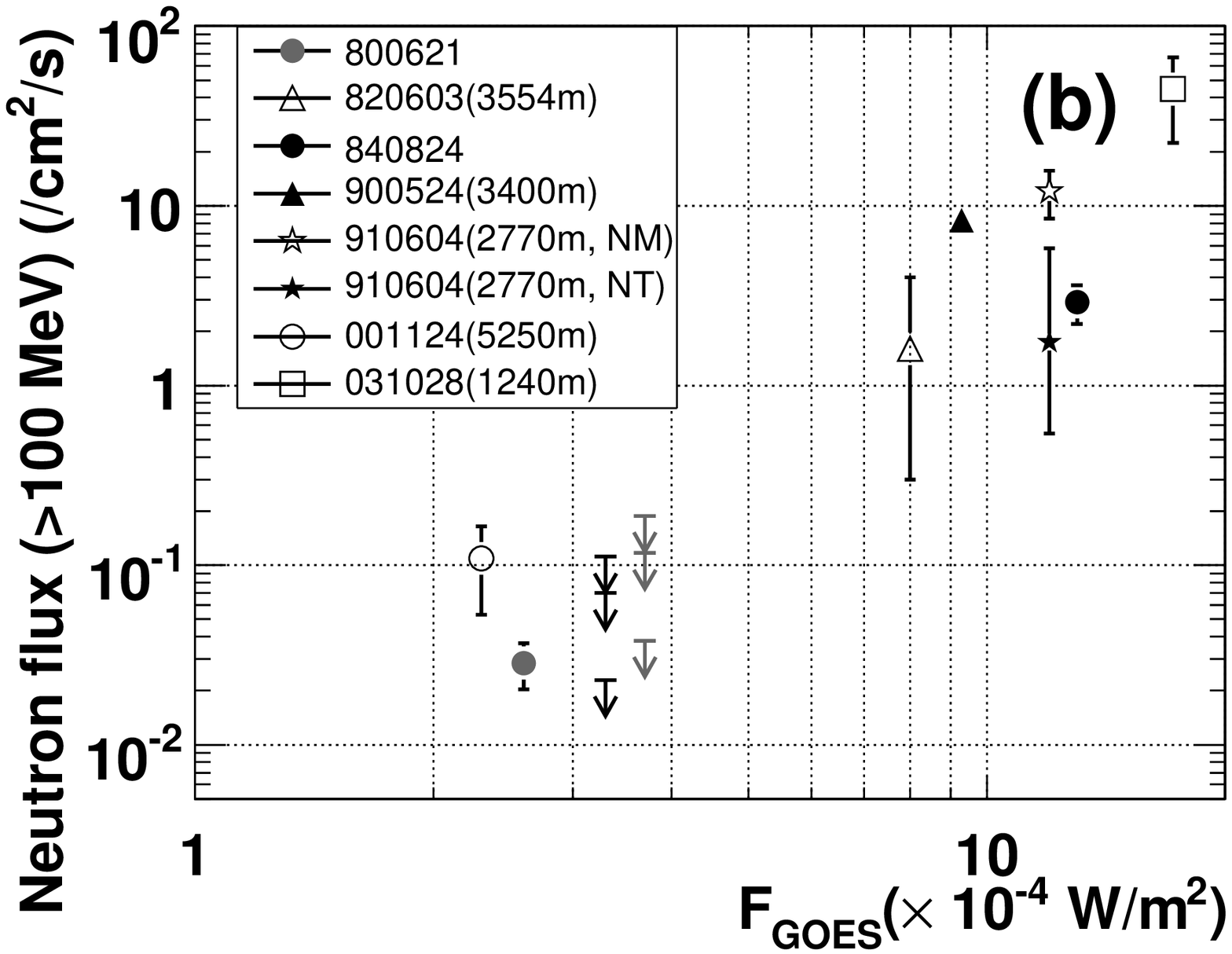}
  \includegraphics[width=8.5cm]{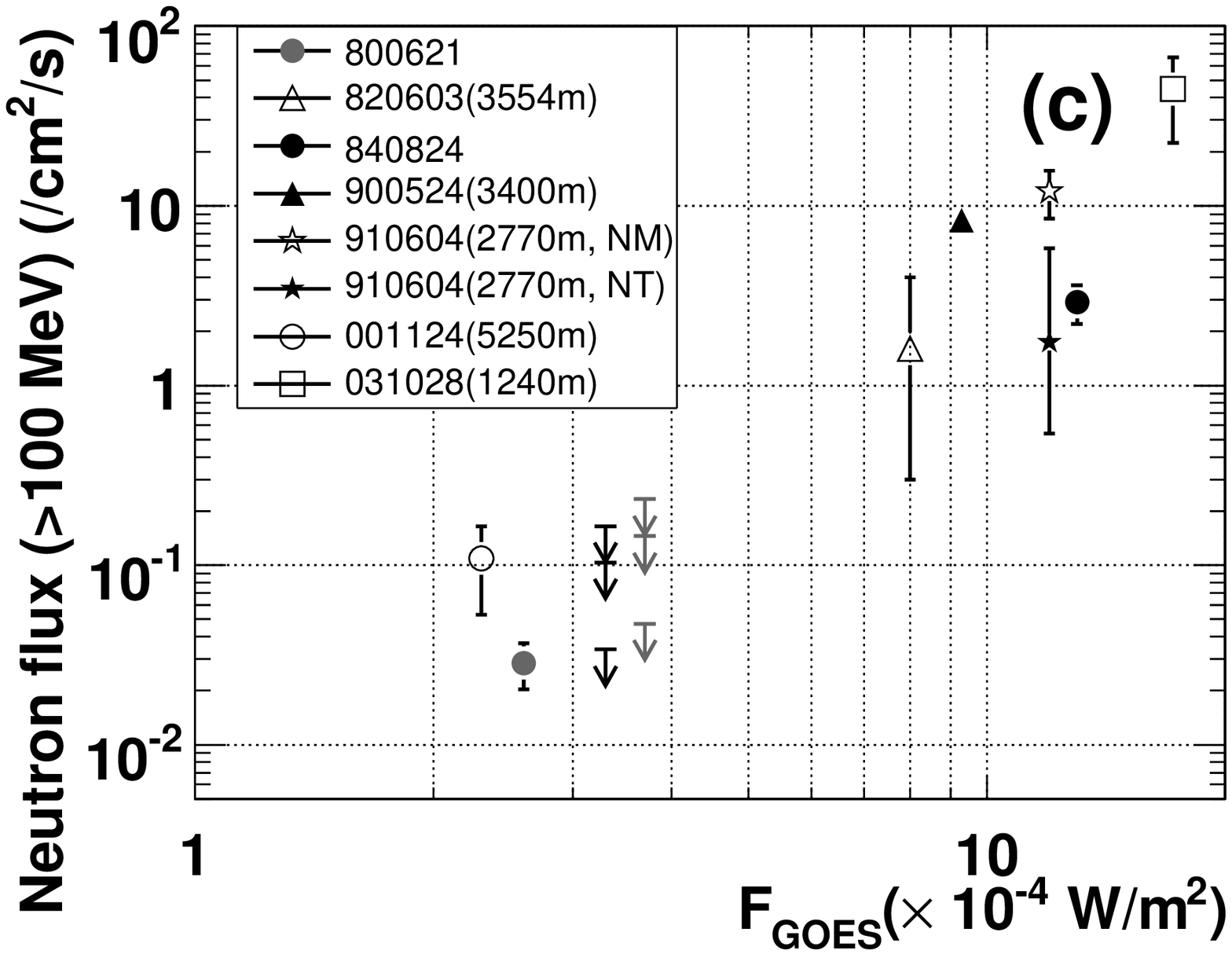}
  \caption{
    The 95 \% CL upper limits on the $>$ 100~MeV solar-neutron flux at the top of the Earth atmosphere, from 
    981122 (gray arrows) and 981128 (black arrows) flares,  are plotted as a function of $F_\mathrm{GOES}$. Previous positive
    detections are also plotted.
    Panels (a), (b), and (c) correspond to $\delta$-, continuous-, and gaussian-emission, respectively.
    Upper limits from low to high correspond to $\beta$ of 3, 4, and 5, respectively. 
    All errors are statistical only.
    Except for 800621 (gray filled circle; \citealt{R1983_800621}) and 840824 flares (black filled circle; \citealt{Eve_840824}), 
    observed by satellites, the station altitude is displayed. The references are ; 
    \citet{Chupp_820603} for the 820603 flare (open triangle),
    \citet{D_900524_1,D_900524_2} for the 900524 flare (filled triangle),
    \citet{W_001124} for the 001124 (open circle), and
    \citet{B_031028} and \citet{W_031028} for the 031028 (open square).
    Neutrons from the 910604 flare were detected by two different detectors installed at Mt. Norikura;
    the neutron monitor~\citep{910604_NM} and the neutron telescope~\citep{910604_NT}.
  }

  \label{fig:ul_X}
\end{figure*}

Figure~\ref{fig:ul_X} and Table~\ref{tab:UL_X} 
show the upper limits on the $>$ 100~MeV solar-neutron flux, $F_{95}$ calculated with equation~(\ref{eq:ul_cal}), for the two (X3.7 and X3.3) of
our final sample flares. For comparison, the previous positive 
detections~\citep{R1983_800621,Chupp_820603,Eve_840824,910604_NT,D_900524_1,910604_NM,D_900524_2,W_001124,B_031028,W_031028}
are also plotted.
Here, the horizontal axis in Fig.~\ref{fig:ul_X} gives the GOES class, and hence 
the values of 1 and 10 correspond to the GOES class of X1 and X10, respectively.
As can be easily seen from Fig.~\ref{fig:ul_X}, 
the present two upper limits are comparable with the lowest neutron flux 
derived from the SMM observation~\citep{Chupp_800621} 
and second lowest one from the highest-altitude NM observation~\citep{W_001124}.
These upper limits, therefore, provide rather stringent constraints on the neutron flux from 
medium X-class flares, 
even though they were observed at large zenith angles of $\sim 50\degr$ (Table~\ref{tab:list_solar_flares}). 

As shown in Fig.~\ref{fig:ul_X}, a positive correlation between absolute neutron fluxes
and GOES class is expected. The derived upper limits are also not inconsistent with such a expected correlation.
Using those absolute fluxes,
Fig.~\ref{fig:ul_X} reveals a relatively tight and steep 
dependence of the solar neutron flux on $F_\mathrm{GOES}$. In fact, the correlation is quantitatively
represented as $F_\mathrm{n} \propto F_\mathrm{GOES}{}^{3.0 \pm 0.2}$ 
(The error of the power-law index is statistical only).
Since the dependence is considerably steeper than direct proportionality, it is strongly suggested that a
larger flare not only accelerates a larger number of nucleons, but also accelerates
them to higher energies. 

SMM observations of 19 X-class flares, ranging from X1 to X20,
have provided a sample of 0.8 -- 7~MeV nuclear gamma-ray fluxes at the top of the Earth atmosphere~\citep{SM1995}. 
Using the sample, we examined in the same way how the nuclear gamma-ray flux depends on $F_\mathrm{GOES}$.
As a result, the dependence was found to be much weaker than that of the neutron flux, quantitatively 
given as $F_{\gamma} \propto F_\mathrm{GOES}{}^{0.32 \pm 0.02}$ (The error is also statistical only).
Thus, the dependence of both secondaries related to ion acceleration on $F_\mathrm{GOES}$ is 
greatly different, at least, within the quoted statistical errors.
The difference between the two relations 
might be in part attributed to a difference in characteristic ion energies 
responsible for these emissions:
nuclear gamma rays are produced primarily by accelerated ions with energies of 1 -- 100~MeV/n, 
while 10 -- 100~MeV/n or even higher-energies are necessary for the accelerated ions to 
efficiently produce high-energy solar neutrons with a 
detectable flux on the ground~\citep{MR1993,ACW2002}.

\begin{figure}
  \centering
  \includegraphics[width=9.0cm]{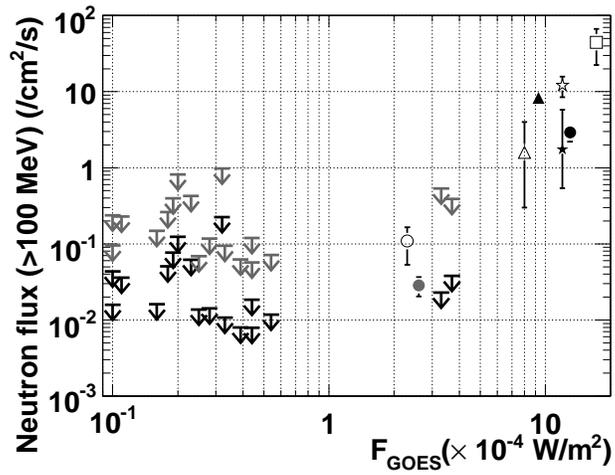}
  \caption{
    The same as Fig.~\ref{fig:ul_X}, but including all the final sample flares.
    Each black arrow shows the most stringent upper limit for each final sample flare, while 
    the least stringent ones are indicated by gray arrows.
    The detail information on each data point are shown in the caption of Fig.~\ref{fig:ul_X}.
  }
  \label{fig:ul_MX}
\end{figure}

In order to further examine the relation between the solar-neutron flux and the flare intensity, 
we show in Fig.~\ref{fig:ul_MX} all upper limits from the final sample flares, and give 
numerical values of the most and the least stringent upper limits in Table~\ref{tab:ul_two_cases}.
The most stringent upper limits in the present work are obtained assuming 
that $\beta$ is 3 and $\nu(t)$ is the continuous-emission, 
while the least stringent ones are derived assuming that $\beta$ is 5 and $\nu(t)$ is the $\delta$-emission. 
From the most stringent upper limits, it is found that the neutron 
flux of M-class flares are probably below 0.01 $\mathrm{cm^{-2}s^{-1}}$ at one AU.
This is consistent with the fact that neither space observatories nor ground-based detectors
have so far succeeded in detecting solar neutrons from M-class solar flares.

\subsection{Dependence on propagation models}\label{sec:model}
Finally, we examine how our upper limits depend on propagation models
used to calculate the atmospheric attenuation factors of solar neutrons.
Presently, several kinds of hadronic interaction models are available to deal with complicated 
nuclear reactions in energies below 10~GeV.
Among them, the present work has employed GHEISHA in CORSIKA 6.500.
However, another model developed by~\citet{Shibata1994} has been widely known as 
one of the representative solar neutron propagation models,
because MC simulations based on the Shibata model well reproduce experimental results 
using 100 -- 400~MeV neutron beams~\citep{RCNP_3,RCNP_1}.
Thus, a comparison between GHEISHA and the Shibata model should be carried out.

Using the 981122 (X3.3) and 981128 (X3.7) flares as representative, 
we computed the flux upper limits using the Shibata model, and compared the results
with those from GHEISHA in Table~\ref{tab:corsika_shibata_flux}.
Thus the upper limits based on the Shibata model are about a quarter of those based on GHEISHA. 
Therefore, this factor should be considered as another systematic uncertainty 
involved in our results.
Since every solar neutron observations on ground is affected by this uncertainty, 
we must use the same propagation model if trying to compare different observations.
\section{Summary and further prospects}
So far, the solar neutron observations and subsequent analyses
have exclusively noticed X-class flares. 
However, to understand ion acceleration in solar flares, 
it is very important to examine how the solar neutron flux depends on the flare intensity.
For this purpose, 
we systematically searched the Yangbajing NM data, taken from 1998 October to 2000 June, for solar neutrons, 
focusing on 166 BATSE-detected M and X-class flares.
No statistically significant signals were found from our small final sample consisting 
of 2 X-class flares and 16 M-class ones.
Hence, the 95\% confidence level upper limits on the $>$ 100~MeV neutron flux were derived for individual flares.
As a consequence, it was found that the upper limits for the two medium X-class flares are comparable
with the neutron flux evaluated from the positive detections by the SMM satellite and the highest-altitude NM.
These results suggest a very steep dependence of the solar neutron flux on the flare size.
Furthermore, we for the first time derived useful upper limits on the solar neutron flux 
from M-class flares.

Some experiments with newly developed high sensitive detectors 
are proposed for space missions~\citep{imaida,FNIT} 
as well as for the ground-based stations~\citep{SSNT}.
However, for the time being, only detectors at very high mountains, like the Yangbajing NM,
would have a potential to provide detections of, or strict constraints on,
the neutron flux from less intense flares, as well as X-class ones. 
We plan to utilize another flare sample from, e.g., the Yohkoh satellite,
and to analyze the NM signals by stacking them in reference to the onset times of a large number
of flares.

\begin{acknowledgements}
  The present research is supported in part by
  the Special Research Project for Basic Science in RIKEN,
  titled ``Investigation of Spontaneously Evolving Systems''. 
  We thank the teams of GOES and BATSE experiments for providing the information on solar X-ray emissions.
  H. T. thanks Dr. K. Watanabe who kindly calculated the attenuation factor of solar neutrons based on the 
  Shibata model.
\end{acknowledgements}

\clearpage
%
\begin{table*}
\begin{minipage}[t]{\columnwidth}
\caption{The 18 solar flares in the final sample}             
\label{tab:list_solar_flares}      
\renewcommand{\footnoterule}{}  
\begin{tabular}{l l c c c c c c c}        
\hline\hline                 
Date & Class/optical importance\footnote{The optical importance of some solar flares is unavailable from the GOES catalog.}  & Position & $t_\mathrm{S}$\footnote{BATSE start time.} & $t_\mathrm{P}$\footnote{BATSE peak time.} & $\Delta t_\mathrm{b}$\footnote{BATSE duration.} & $N_\mathrm{P}$\footnote{BATSE peak rate.} & $N_\mathrm{T}$\footnote{BATSE total counts.}&$\theta_\mathrm{SUN}$\footnote{A zenith angle of the Sun for individual flares at the onset time.}  \\    
(YYMMDD) & {}     & {}  & (UT)             & (UT)            & (sec)          & ($2000\mathrm{cm}^{-2}\mathrm{s}^{-1}$) & {} & (deg.) \\
\hline                        
981112    &   M1.0/1N   &  N21W34 & $ 5^{h}26^{m}23^{s} $      &    $ 5^{h}27^{m}6^{s} $  & 124   & 45542   & 777267 & 48 \\ 
981122    &   X3.7/1N   &  S27W82 & $ 6^{h}33^{m}36^{s} $      &    $ 6^{h}39^{m}17^{s} $  & 901  & 1130000 & 168000000 & 51 \\ 
981128    &   X3.3/3N   &  N17E32 & $ 5^{h}31^{m}36^{s} $      &    $ 5^{h}40^{m}46^{s} $  & 2834 & 670736  & 120056656 & 52 \\ 
981217    &   M3.2/1N   &  S27W46 & $ 7^{h}41^{m}41^{s} $      &    $ 7^{h}43^{m}16^{s} $  & 136  & 120179  & 3418773   & 59 \\ 
990402    &   M1.1/     &  {}     & $ 8^{h}9^{m}14^{s} $      &    $ 8^{h}11^{m}48^{s} $  & 453   & 5960    & 727034    & 38 \\ 
990404    &   M5.4/1F   &  N18E72 & $ 5^{h}24^{m}57^{s} $      &    $ 5^{h}25^{m}9^{s} $  & 114   & 1085    & 63400     & 26 \\ 
990503    &   M4.4/2N   &  N15E32 & $ 5^{h}39^{m}55^{s} $      &    $ 5^{h}47^{m}52^{s} $  & 2523 & 26942   & 6403993   & 14 \\ 
990510    &   M2.5/2N   &  N16E19 & $ 5^{h}24^{m}35^{s} $      &    $ 5^{h}26^{m}45^{s} $  & 561  & 13420   & 955186    & 14 \\ 
990529    &   M1.6/     &  {}     & $ 3^{h}6^{m}43^{s} $      &    $ 3^{h}8^{m}50^{s} $  & 885    & 17123   & 2420462   & 39 \\ 
990604    &   M3.9/2B   &  N17W69 & $ 6^{h}53^{m}29^{s} $      &    $ 6^{h}59^{m}58^{s} $  & 1076 & 490858  & 44716592  & 14 \\ 
990724    &   M3.3/SF   &  S28E78 & $ 7^{h}49^{m}54^{s} $      &    $ 8^{h}1^{m}24^{s} $  & 849   & 19173   & 1727533   & 25 \\ 
990725    &   M1.0/1F   &  S26E58 & $ 4^{h}22^{m}17^{s} $      &    $ 4^{h}24^{m}1^{s} $  & 133   & 806     & 13297     & 27 \\ 
991116a   &   M1.8/SF   &  N18E43 & $ 4^{h}48^{m}0^{s} $      &    $ 4^{h}48^{m}52^{s} $  & 105   & 651     & 32910     & 50 \\ 
991116b   &   M2.3/1N   &  S14E10 & $ 6^{h}3^{m}38^{s} $      &    $ 6^{h}7^{m}31^{s} $  & 491    & 21992   & 2695715   & 49 \\ 
991126    &   M1.9/2B   &  S19E58 & $ 3^{h}52^{m}56^{s} $      &    $ 3^{h}55^{m}8^{s} $  & 665   & 18033   & 1763834   & 58 \\ 
000408    &   M2.0/1B   &  S15E26 & $ 2^{h}38^{m}13^{s} $      &    $ 2^{h}38^{m}38^{s} $  & 523  & 189878  & 7243484   & 52 \\ 
000504    &   M2.8/1N   &  S14W90 & $ 4^{h}34^{m}33^{s} $      &    $ 4^{h}37^{m}28^{s} $  & 744  & 34745   & 3111922   & 23 \\ 
000515    &   M4.4/     &    {}   & $ 8^{h}33^{m}43^{s} $      &    $ 8^{h}42^{m}46^{s} $  & 3566 & 38435   & 18834356  & 33 \\ 
\hline                                   
\end{tabular}
\end{minipage}
\end{table*}
\clearpage
\begin{table*}
\begin{minipage}[t]{\columnwidth}
\caption{The obtained $N_\mathrm{S}$, $\Delta N_\mathrm{S}$, $N_{95}$, and $\sigma_\mathrm{S}$ for each neutron emission profile.}             
\label{tab:Ns_count}      
\renewcommand{\footnoterule}{}  
\begin{tabular}{l c c c c c c c c c c c c}        
\hline\hline
Date    & \multicolumn{4}{c}{$\delta$-emission} &   \multicolumn{4}{c}{continuous-emission} & \multicolumn{4}{c}{gaussian-emission} \\
(YYMMDD) &  $N_\mathrm{S}$ & $\Delta N_\mathrm{S}$ & $N_\mathrm{95} $ & $\sigma_\mathrm{S}$ 
         & $N_\mathrm{S}$  & $\Delta N_\mathrm{S}$ & $N_\mathrm{95} $ & $\sigma_\mathrm{S}$ 
         & $N_\mathrm{S}$  & $\Delta N_\mathrm{S}$ & $N_\mathrm{95} $ & $\sigma_\mathrm{S}$  \\
\hline                        
981112 & -3700 & 3719 & 5263 & -1.0 & -4424 & 4221 & 5874 & -1.1 & -4182 & 4002 & 5572 & -1.1 \\ 
981122 & -756 & 3675 & 6724 & -0.2 & -8173 & 6037 & 7644 & -1.4 & -4545 & 5470 & 8152 & -0.8 \\ 
981128 & 1418 & 3795 & 8434 & 0.4 & -18620 & 8989 & 9273 & -2.1 & -11409 & 6264 & 6926 & -1.8 \\ 
981217 & 3646 & 3532 & 9734 & 1.0 & 4371 & 4175 & 11558 & 1.1 & 3746 & 3886 & 10480 & 1.0 \\ 
990402 & 2896 & 3533 & 9098 & 0.8 & 4420 & 4979 & 13105 & 0.9 & 4479 & 4449 & 12159 & 1.0 \\ 
990404 & -4626 & 3754 & 4934 & -1.2 & -5345 & 4092 & 5261 & -1.3 & -4776 & 3800 & 4957 & -1.3 \\ 
990503 & -1833 & 3603 & 5962 & -0.5 & 4974 & 8601 & 20464 & 0.6 & -243 & 6056 & 11716 & 0.0 \\ 
990510 & 656 & 3475 & 7259 & 0.2 & 4582 & 5259 & 13767 & 0.9 & 2229 & 4231 & 9896 & 0.5 \\ 
990529 & -1769 & 3458 & 5716 & -0.5 & -6847 & 5862 & 7856 & -1.2 & -2857 & 4197 & 6564 & -0.7 \\ 
990604 & -478 & 3488 & 6530 & -0.1 & -1514 & 6251 & 11303 & -0.2 & 2274 & 5583 & 12550 & 0.4 \\ 
990724 & -54 & 3495 & 6817 & 0.0 & -3077 & 5892 & 9704 & -0.5 & 881 & 4431 & 9288 & 0.2 \\ 
990725 & -608 & 3486 & 6445 & -0.2 & -1479 & 4115 & 7154 & -0.4 & -581 & 3735 & 6950 & -0.2 \\ 
991116 & -4199 & 3536 & 4712 & -1.2 & -4866 & 3978 & 5241 & -1.2 & -3539 & 3901 & 5676 & -0.9 \\ 
991116 & 1794 & 3735 & 8602 & 0.5 & 1516 & 5040 & 10933 & 0.3 & 2032 & 4780 & 10808 & 0.4 \\ 
991126 & -4786 & 3447 & 4322 & -1.4 & -5658 & 5336 & 7396 & -1.1 & -4990 & 4207 & 5609 & -1.2 \\ 
000408 & 6955 & 3635 & 12986 & 1.9 & 8412 & 5188 & 17086 & 1.6 & 6986 & 3756 & 13224 & 1.9 \\ 
000504 & 3205 & 3589 & 9462 & 0.9 & 2791 & 5628 & 13027 & 0.5 & 2276 & 4694 & 10828 & 0.5 \\ 
000515 & -2345 & 3726 & 5927 & -0.6 & 16249 & 9783 & 32585 & 1.7 & 1746 & 6108 & 13183 & 0.3 \\ 
\hline
\end{tabular}
\end{minipage}
\end{table*}
\clearpage
\begin{table}
\begin{minipage}[t]{\columnwidth}
\caption{The 95 \% CL upper limits on the $>$ 100~MeV solar-neutron flux for two X-class flares.}             
\label{tab:UL_X}      
\renewcommand{\footnoterule}{}  
\begin{tabular}{c c c c c c c}        
\hline\hline                 
 {}       & \multicolumn{3}{c}{981128 (X3.3)} &   \multicolumn{3}{c}{981122 (X3.7)} \\
 {}       & \multicolumn{3}{c}{$(\times 10^{-2}$ $\mathrm{cm^{-2}s^{-1}})$}  & \multicolumn{3}{c}{$(\times 10^{-2}$ $\mathrm{cm^{-2}s^{-1}})$} \\ \hline
$\beta$~\footnote{The power-law index}    &  $\delta$.E.\footnote{$\delta$-emission.}    &   G.E.~\footnote{gaussian-emission.}    & C.E.\footnote{continuous-emission.}        &  $ \delta$.E.    &   G.E.  &  C.E.   \\
3 & 10.9 & 3.4 & 2.3 & 7.9 & 4.7 & 3.8 \\ 
4 & 33.4 & 10.3 & 7.0 & 24.2 & 14.5 & 11.7 \\ 
5 & 53.2 & 16.5 & 11.2 & 38.9 & 23.3 & 18.8 \\ 
\hline
\end{tabular}
\end{minipage}
\end{table}
\clearpage
\begin{table} 
\begin{minipage}[t]{\columnwidth}
\caption{The most and least stringent upper limits on the $>$ 100~MeV solar-neutron flux.}             
\label{tab:ul_two_cases}      
\renewcommand{\footnoterule}{}  
\begin{tabular}{l c c c c c c}        
\hline\hline                 
Date     & Class/optical imp. & $\mathrm{UL_{M}}$\footnote{The most stringent upper limit.} & $\mathrm{UL_{L}}$ \footnote{The least stringent upper limit.} \\
(YYMMDD) & {} & $(\times 10^{-2}$ $\mathrm{cm^{-2}s^{-1}})$ & $(\times 10^{-2}$ $\mathrm{cm^{-2}s^{-1}})$  \\
\hline
981112 & M1.0/1N&4.4 & 23.8 \\
981122 & X3.7/1N&3.8 & 38.9 \\
981128 & X3.3/3N &2.3 & 53.2 \\
981217 & M3.2/1N &22.5 & 97.5 \\
990402 & M1.1/&3.6 & 22.8 \\
990404 & M5.4/1F&1.2 & 7.1 \\
990503 & M4.4/2N&0.8 & 5.7 \\
990510 & M2.5/2N &1.4 & 6.9 \\
990529 & M1.6/ &1.6 & 14.8 \\
990604 & M3.9/2B&0.8 & 6.2 \\
990724 & M3.3/SF &1.1 & 9.4 \\
990725 & M1.0/1F&1.6 & 9.5 \\
991116a& M1.8/SF&5.1 & 26.2 \\
991116b& M2.3/1N&6.2 & 42.8 \\
991126 & M1.9/2B&7.7 & 39.6 \\
000408 & M2.0/1B&12.4 & 82.0 \\
000504 & M2.8/1N&1.4 & 11.8 \\
000515 & M4.4/&1.8 & 12.0 \\
\hline                                   
\end{tabular}
\end{minipage}
\end{table}
\clearpage
\begin{table*}
\begin{minipage}[t]{\columnwidth}
\caption{A comparison between upper limits based on two propagation models.}
\label{tab:corsika_shibata_flux}      
\renewcommand{\footnoterule}{}  
\begin{tabular}{l c c c c c c c} 
\hline\hline                 
{}          &     {}  & \multicolumn{3}{c}{981122}           & \multicolumn{3}{c}{981128} \\
emission    &     {}  &  Shibata       & GHEISHA   & {}   & Shibata       & GHEISHA   &  {} \\
profile     & $\beta$ & \multicolumn{2}{c}{($\times 10^{-2}$ $\mathrm{cm^{-2}s^{-1}}$)} & Ratio\footnote{A ratio of the Shibata model to GHEISHA.}
                      & \multicolumn{2}{c}{($\times 10^{-2}$ $\mathrm{cm^{-2}s^{-1}}$)} & Ratio \\
\hline
{}          &  3      &  2.0    & 7.9    & 0.26 & 2.6 & 10.9 & 0.23 \\
$\delta$-emission &  4      &  6.5    & 24.2   & 0.27 & 8.1  & 33.4 & 0.24  \\
{}          &  5      &  9.9    & 38.9   & 0.26 & 12.5 & 53.2 & 0.23 \\
\hline
{}          &  3      &  1.2    & 4.7    & 0.26 & 0.8  & 3.4 & 0.23  \\
gauss-emission        &  4      &  3.9    & 14.5   & 0.27 & 2.5 & 10.3 & 0.24   \\
{}          &  5      &  6.0    & 23.3   & 0.26 & 3.9 & 16.5 & 0.23 \\
\hline
{}          &  3      &  1.0    & 3.8    & 0.26 & 0.5 & 2.3 & 0.23 \\
continuous-emission        &  4      &  3.1    & 11.7   & 0.27 & 1.7 & 7.0 & 0.24\\
{}          &  5      & 4.8     & 18.8   & 0.26 & 2.6 & 11.2 & 0.23\\
\hline
\end{tabular}
\end{minipage}
\end{table*}
\clearpage
%
%
%
%
%
\clearpage
\clearpage
\clearpage
%
%
\clearpage
%
%
%
\clearpage
%
%
%
\clearpage
\clearpage
\bibliographystyle{aa}
\end{document}